\documentclass[11pt,twocolappendix]{emulateapj}

\pdfoutput=1

\usepackage{graphicx}
\usepackage{psfig}
\usepackage{float}

\usepackage[sort&compress]{natbib}

\usepackage{subfigure}
\newcommand{\degree}{\ensuremath{^\circ \,}}

\shorttitle{Modeling Optically Thick Comets}
\shortauthors{Gersch \& A'Hearn}
\submitted{Submitted to ApJ} 

\begin{document}

\nocite{*}

\title{Coupled Escape Probability for an Asymmetric Spherical Case: Modeling Optically Thick Comets}

\author{Alan M. Gersch}
\affil{Department of Astronomy, University of Maryland, College Park, MD 20742-2421 }
\email{agersch@astro.umd.edu}

\author{Michael F. A'Hearn}
\affil{Department of Astronomy, University of Maryland, College Park, MD 20742-2421 }

\begin{abstract}
 We have adapted  Coupled Escape Probability, a new exact method of solving radiative transfer problems, for use in asymmetrical spherical situations. Our model is intended specifically for use in modeling optically thick cometary comae, although not limited to such use. This method enables the accurate modeling of comets' spectra even in the potentially optically thick regions nearest the nucleus, such as those seen in Deep Impact observations of 9P/Tempel 1 and EPOXI observations of 103P/Hartley 2.
\end{abstract}


\section{Introduction}

Comets are understood to be frozen remnants from the formation of our solar system. As such, their chemical composition is of great significance to understanding the origin of the planets and the distribution of important molecules, including water, throughout the solar system.
This was and is a major goal of the Deep Impact and EPOXI Missions, among others, as well as ground-based observations of comets.

Recent observations, in particular those of the Deep Impact and EPOXI Missions (see e.g. Feaga, et al. 2007 or A'Hearn, et al. 2011), have provided better spectra of a cometary coma than were, in general, previously available. These observations include spectra with high spatial resolution very near to the nucleus. 
No previous observations had as much spatially well-resolved spectral data, and thus there was little observationally driven need to pay special or close attention to the densest part of the coma. Ground-based observations could only see optically thick regions of comae for the brightest and/or most active of comets. (e.g. Hale Bopp; see DiSanti et al. 2001.)

Therefore many earlier studies that modeled  spectra of comae, in keeping with the available observations of the time, did not attempt to calculate optical depth effects on spectra. Optically thin comae were assumed, since the field of view in those observations being modeled would be dominated by the majority of the coma far from the nucleus, which is optically thin (see e.g. Chin \& Weaver 1984; Crovisier \& Le Bourlot 1983).
However, with the proliferation of space missions to comets, as well as much better instruments for ground-based observations (see, e.g. DiSanti et al. 1999), this is no longer a truly tenable approach.

Our goal is to better understand the abundances, distributions and creation mechanisms of various gases observed in comae, in particular of comet 9P/Tempel 1, the target of the Deep Impact Mission, and 103P/Hartley 2, the subject of the EPOXI mission.

In order to do so, we have built a computer model of the spectrum of the comet's coma which includes the difficult and often ignored problem of accurately including radiative transfer to account for the potentially optically thick coma (or regions of the coma) near the nucleus.

This model will facilitate analyzing the actual spectral data from the Deep Impact and EPOXI missions to better determine abundances of key species, including CO, CO$_{2}$, and H$_{2}$O, as well as remote sensing data on active comets.

\subsection{The Simple Coma Model}

We begin our modeling of IR ro-vibrational spectra of a coma by initially following the method used by Chin \& Weaver 1984, Crovisier 1987, and others with some minor improvements. 
As in those models, we assume a constant expansion velocity, thus linearly relating any radial distance to a specific time since a ``parcel'' of gas was released from the surface of the nucleus. Therefore we numerically integrate over time the linear differential equations defined by the Einstein coefficients and collisional rate coefficients to get fractional molecular enegry level populations for each distance from the nucleus, from which we calculate emission spectra. 

\begin{eqnarray}
\label{bigEqn}
\frac{dn_{k}}{dt} = \sum_{i} ( A_{ik} n_{i} + J_{\nu_{ik}} ( B_{ik} n_{i} - B_{ki} n_{k} ) ) + \\
\nonumber        - \sum_{l} ( A_{kl} n_{k} + J_{\nu_{kl}} ( B_{kl} n_{k} - B_{lk} n_{l} ) ) + \\
\nonumber	 \sum_{j} C_{jk} ( n_{k} - n_{j} e^{-E_{jk}/k_{B} T} )
\end{eqnarray}
Here $i,j, k$ and $l$ indicate energy levels of a molecule with the $n$'s with those indices being the corresponding level populations. The Einstein coefficients between levels x and y are $A_{xy}$ and $B_{xy}$, and C is a similar collisional coefficient. We use $J_{\nu_{xy}}$ for the mean intensity of radiation at the frequency corresponding to the transition between x and y, $E_{xy}$ is the energy difference between levels x and y, and $k_{B}$ and T have their usual meanings. The summations are over all levels i, l or j which have a transition into or out of level k. For collisional coefficients $C = n_{_{H_{2}O}} \sigma \bar{v}$ where $n_{_{H_{2}O}}$ is the number density of H$_{2}$O (assumed to be the dominant collisional partner) $\sigma$ is the collisional cross section for a given transition and $\bar{v}$ is the mean (thermal) molecular speed. 
This allows us to include a time variable production rate. We use such a coma integration as the initial basis for our coma  model before including potential optical depth effects.

Our primary improvement is the inclusion of radiative transfer calculations using our spherical adaptation of the Coupled Escape Probability method (hereafter, ``CEP''; see Elitzur \& Asensio Ramos 2006, hereafter, ``CEP06'') to more correctly model optically thick (or potentially thick) regions of cometary comae. This is described in detail below, and is the main part of this paper.
We use the coma integration results to provide the ``initial guess'' values for populations used in the subsequent radiative transfer calculations using CEP.

\begin{figure}[h]
\centerline{\includegraphics[width=0.2\textwidth,height=3in,angle = 90]{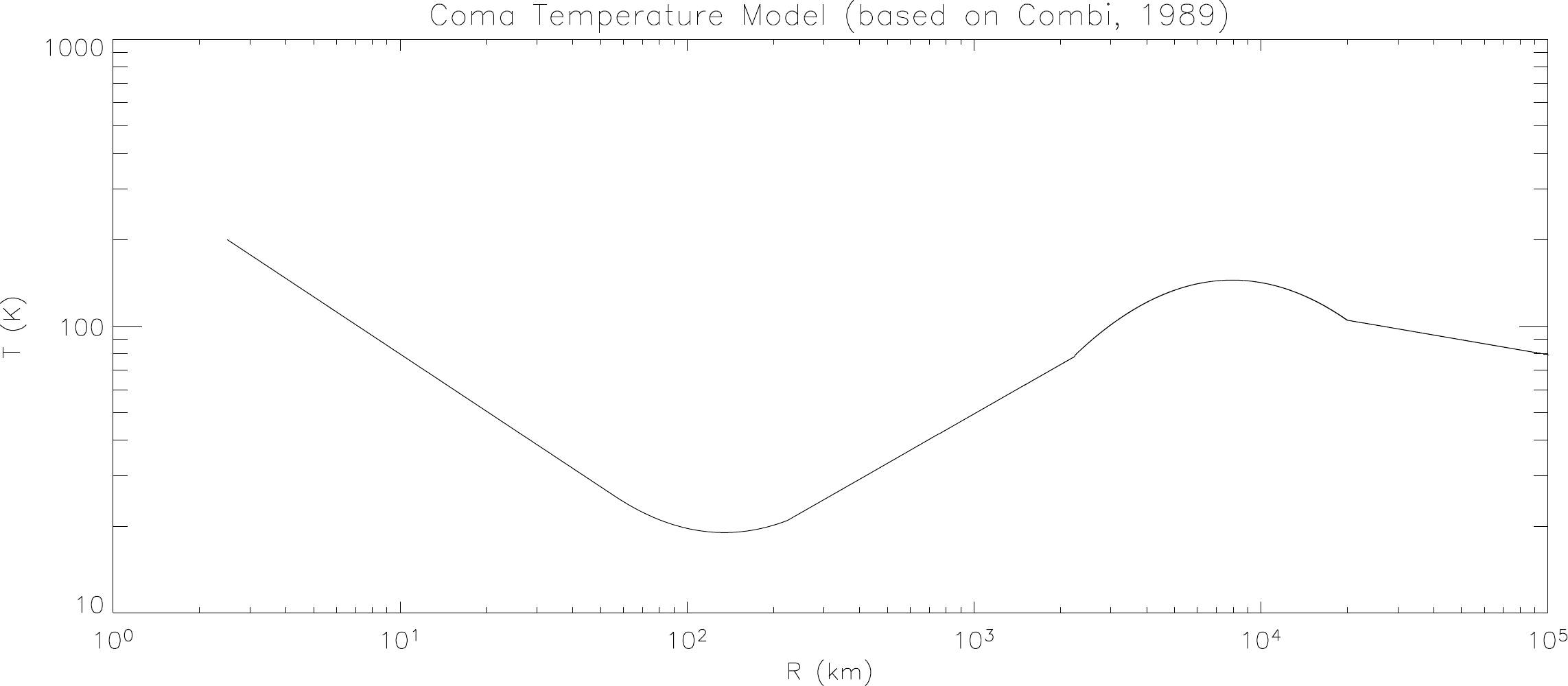}}
\caption{Temperature profile of cometary atmosphere based on Combi 1989 for surface temperature  of 200 K and radial gas velocity of 0.8 km s$^{-1}$. \label{combi200} }
\end{figure}

For the purposes of the initial coma model, we treat the comet as spherically symmetric, and as having a uniform and constant gas production rate over its entire surface. The outward speed of the gas is also assumed to be constant, as per Chin \& Weaver 1984. While this is not strictly physically accurate (see e.g. Combi 1996) the variation over the majority of the coma is relatively small. We use a temperature profile that varies with radial distance from the nucleus, having closely followed Combi's 1989 model (see Fig. \ref{combi200}.) These approximations make integration over time equivalent to calculating these values over increasing distances from the comet nucleus for a ``shell'' of gas expanding outwards from the nucleus.

We ignore the photodestruction of CO in our coma model, due to its long lifetime (see Crovisier 1994.) The lifetime is $>10^7$ s, and we are integrating out to $10^5$ km with an expansion velocity of 0.8 km s$^{-1}$. (Note that others, e.g. Morgenthaler, et al. 2011, find a shorter lifetime, but still $\gtrsim 4 \times 10^5$ s, which is still large enough that it can be neglected in our modeling out to 10$^{5}$ km.)

The model can include coma morphology features as well, each modeled with its own coma integration using separate conditions. Such features, as seen in the Deep Impact and EPOXI encounters, are a main motivation for creating this model to better understand possible optical depth effects in the near nucleus regions of the coma. (See e.g. Feaga, et al. 2007 or A'Hearn, et al. 2011)

After describing our method in Section 2, we present (in Section 3) some results demonstrating its use in better understanding possible optically thick spectra for the carbon monoxide 1-0 (X$^{1} \Sigma^{+}$) band in spherically symmetric comae. These may be useful for ground-based observers (or space telescope observations of comets) to better fathom the depths of cometary spectra. Forthcoming model results for CO$_{2}$, H$_{2}$O and near-nucleus morphological features will follow (in other papers).

\section{Our Method: CEP Adapted for  an Asymmetric Spherical Case}

This section describes our adaption of the Coupled Escape Probability  radiative transfer technique to spherical cases in which the plane parallel approximation is not appropriate, including most cometary problems. (Note that Yun, et al. 2009, previously adapted CEP for {\it purely symmetric} spherical cases.) We have developed this model specifically for use in studying cometary comae, but it could be applied to many other astrophysical phenomena as well (e.g. planetary atmospheres, molecular clouds, etc.).

We derive expressions analogous to the relevant CEP06 equations, adapted from the plane parallel situation to spherically symmetric cases. Then we include asymmetries of two different types: radiation from an outside source (here, the Sun) and non-uniform conditions due to morphology.

In brief, the CEP method (see CEP06) divides up a plane-parallel ``slab'' into ``zones'' (each of which has uniform properties) and calculates the net radiative bracket (``$p$'') that is multiplied by the Einstein A coefficient in the equations of statistical equilibrium (see Equation \ref{bigEqn}) for each radiative transition for each zone. (Note that the ``p'' term effectively combines the ``B'' terms into the ``A'' term. See CEP06 for more detail.) The innovation of CEP is that the net radiative brackets for each zone can accurately represent the contributions of {\it all} zones' emission and absorption to/from other zones. (As opposed to ``plain'' Escape Probability where a similar factor added to the statistical equilibrium equations is only a {\it local} approximation of a photon's likelihood of escaping the entire slab. See, Bockelee-Morvan 1987, Zakharov, et al. 2007 and  Litvak \& Kuiper 1982.) The statistical equilibrium equations for all zones, with the inclusion of the  net radiative bracket, form a single non-linear matrix. This matrix can be solved using an algorithm for non-linear matrix solving such as Newton's Method. We use functions from Numerical Recipes in C (see Press, et al. 1992) for the Newton based matrix solver, as well as other calculations such as numerical integration. Further computational details can be found in the Appendix. This solution yields the fractional populations of molecular energy levels for each zone. From these, the flux emitted by the slab can be calculated. See CEP06 for more details and the derivations of the original plane-parallel equations to which we make reference below.

\subsection{Theoretical/Analytical Expressions for the Net Radiative Bracket}

In CEP06 equation 7, Elitzur \& Asensio Ramos derive a purely analytical expression for the net radiative bracket in a plane parallel slab, based on the formal solution of the radiative transfer equation and the definition of the net radiative bracket: 

$$p(\tau) =  $$
\begin{equation}
\label{CEPEqn7}
1 - \frac{1}{2 S(\tau)} \int_{0}^{\tau_{t}} S(t) dt \int_{-\infty}^{\infty} \Phi^2(x) dx  \int_{0}^{1} e^{-|\tau-t|\Phi(x)/\mu} \frac{d\mu}{\mu}
\end{equation}
where $\tau$ and $t$ refer to optical depths for a specific wavenumber, with $\tau_{t}$ being the overall optical thickness of the slab, $S(\tau)$ or $S(t)$ to the source function for a given line, ($S_{\nu_{21}} = \frac{A_{21} n_{2}}{B_{12} n_{1} -B_{21} n_{2}}$) $\Phi(x)$ is a dimensionless line profile, and $\mu =$ cos$\theta$, where $\theta$ is the angle of a given ray measured from the normal to the plane.

We use a spherical analog to this theoretical expression (i.e. as opposed to the discretized expression they introduce later involving a number of ``zones'') for the net radiative bracket:

$$ p( \tau(r,\theta,\phi) ) =$$
\begin{eqnarray}
\label{analyticalEqn}
1 - \int_{4\pi} \left [ \frac{1}{2 S( \tau(r,\theta,\phi, \Omega) )}  \int_{0}^{\tau(r,\theta,\phi, \Omega)}  S( t(r,\theta,\phi, \Omega) \;\;\; \right.\nonumber \\
 \left.  \times \; \int_{-\infty}^{\infty} \Phi^2(x)  \left ( e^{-|\tau(r,\theta,\phi, \Omega)-t(r,\theta,\phi, \Omega)|\Phi(x)} \right ) dx \; dt   \right ] d\Omega \;\;\;\;\;\;
\end{eqnarray}

where $p( \tau(r,\theta,\phi) )$ refers to the net radiative bracket at any point labeled by the coordinates $(r,\theta,\phi)$ in spherical coordinates with the origin at the center of some sphere of interest. (In our study, the sphere is centered on the comet nucleus; it could be the center of any spherical astronomical object for any given case.)

As in CEP06,  $\Phi(x) = (\pi \, \Delta\nu_{D})^{-1/2}\, e^{-x^{2}}$ is a dimensionless line profile, normalized so that  $\int \Phi(x) dx = 1$, where $x = (\nu - \nu_0)/ \Delta\nu_{D}$ is the dimensionless line shift from line center $\nu_0$ and $\Delta\nu_{D}$ is the doppler line width, $\nu_0/c ( 2kT/m)^{1/2}$.
In the present work, we neglect doppler offsets between different locations in the coma with relative velocities. This is due to our work being primarily motivated by the Deep Impact observations of the near-nucleus coma (see Feaga, et al. 2007), for which gas tends to be released, broadly speaking, towards one direction/side of the nucleus. For this situation, doppler shifts are likely to be less than line broadening. In future work, we plan to include a more generally accurate treatment.

\begin{figure}[h]
\centerline{\includegraphics[trim=15cm 17.5cm 15cm 2cm, clip=true, width=0.5\textwidth]{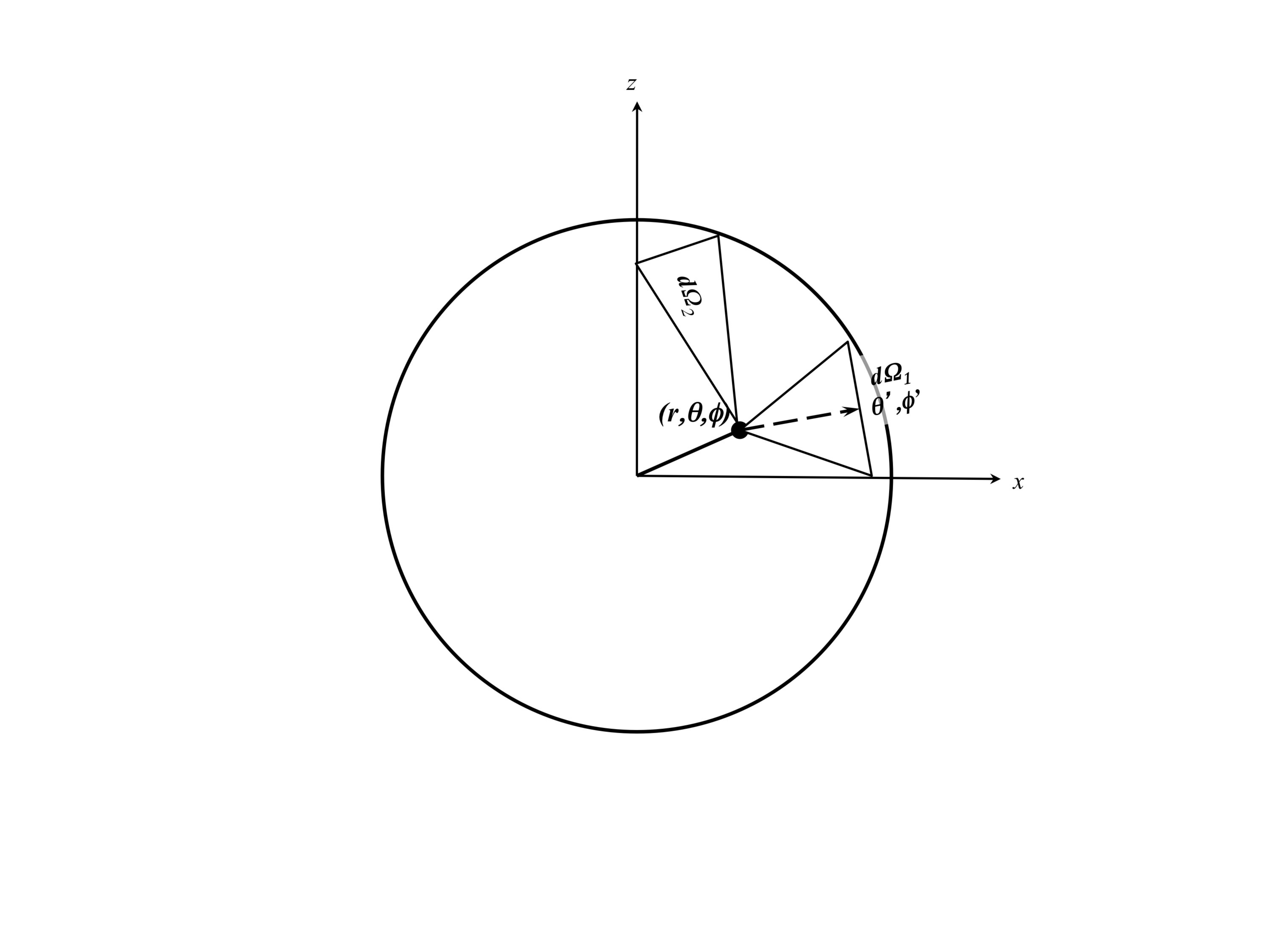}}
\caption{Sphere showing a point $(r,\theta,\phi)$ and two solid angle elements $d\Omega_1$ and $d\Omega_2$. (Viewed from the $-y$ direction.) \label{fig1}}
\end{figure}

Both $\tau(r,\theta,\phi, d\Omega)$ and $t(r,\theta,\phi, d\Omega)$ refer to optical depths and $S( \tau(r,\theta,\phi, d\Omega) )$ to the source function {\it as viewed from the coordinates} $(r,\theta,\phi)$  along a ``pencil'' of solid angle $d\Omega$. (See Fig. \ref{fig1}.)

The optical depth along a ``pencil'' (or cone) of solid angle $d\Omega$ is, of course, highly dependent on the particular direction of a given solid angle element $d\Omega$, i.e. dependent on $\theta ',\phi '$, the direction of a vector centered on $(r,\theta,\phi )$ pointing along the centerline of $d\Omega$. (See Fig. \ref{fig1}.)

\subsection{Discrete Expressions for $p$}

The above CEP06 expression for $p(\tau)$ (Equation \ref{CEPEqn7}) is turned into a discrete expression by dividing the slab into multiple ``zones'' (labeled with indices $i$ \& $j$). This yields CEP06 equation 14:
\begin{eqnarray}
\label{CEPEqn14}
p^{i} = 1 - \frac{1}{2 \tau^{i,i-1} S^{i}} \sum_{j=1}^{z} S^{j} \times \int_{\tau^{i-1}}^{\tau^{i}} d\tau \int_{\tau^{j-1}}^{\tau^{j}} dt  \nonumber \\*
\times \int_{-\infty}^{\infty} \Phi^2(x) dx \int_{0}^{1} e^{-|\tau-t|\Phi(x)/\mu}  \frac{d\mu}{\mu} \nonumber \\
\end{eqnarray}

This expression for $p$ can be calculated for each zone (and each wavenumber/frequency) and the resultant $p$'s included in the equations of statistical equilibrium, which are then solved. (See also CEP06 equations 32-36, and accompanying CEP06 text, for a more complete discussion.)

\begin{figure}[h]
\centerline{\includegraphics[width=0.5\textwidth,trim=15cm 17.5cm 15cm 2cm, clip=true]{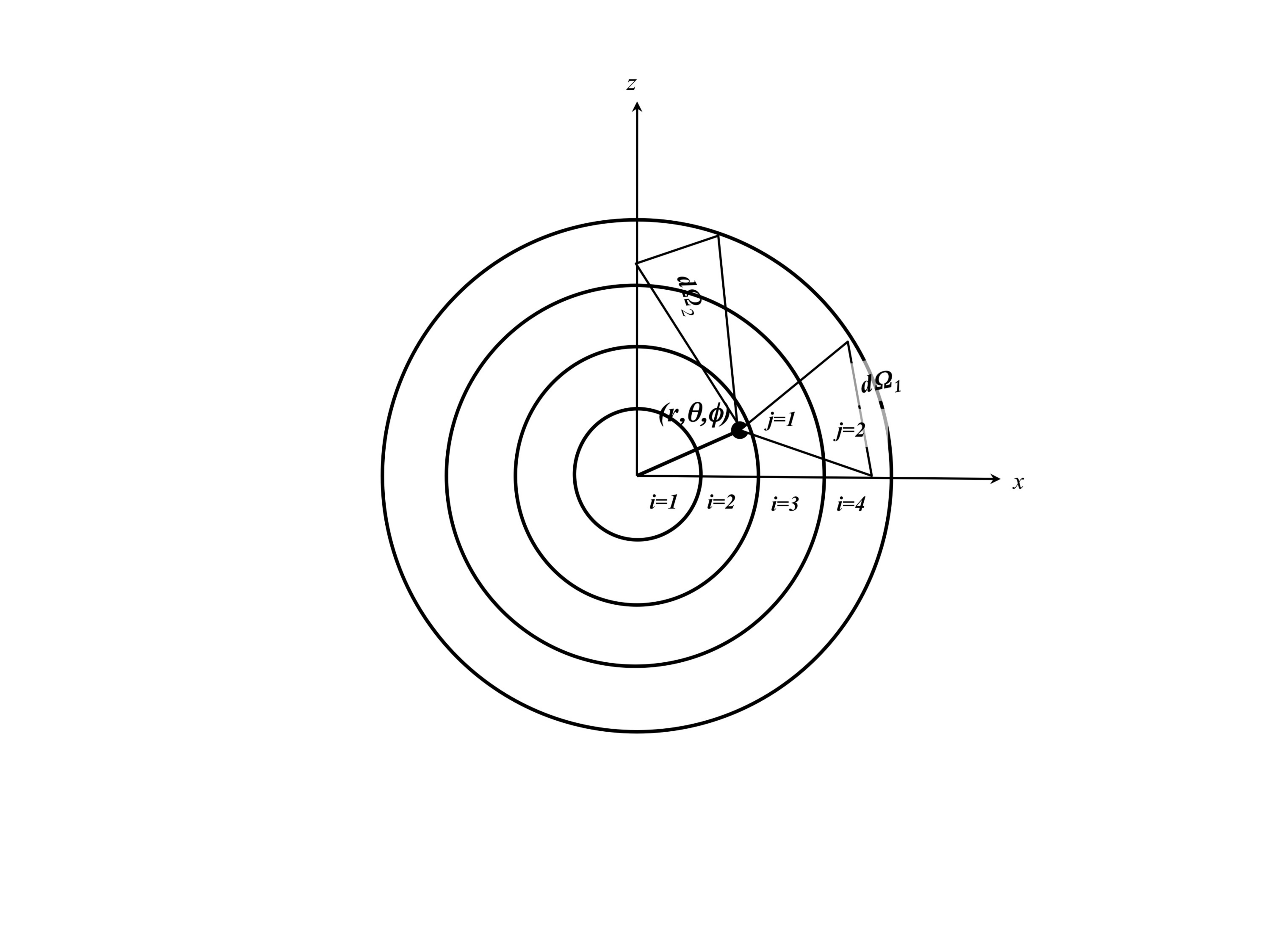}}
\caption{Example sphere showing division into 4 shells, $i=1..4$ and solid angle $d\Omega_1$ for which $z=2$ with indices $j=1,2$, as per Equation \ref{partlyDiscreteEqn} (Viewed from the $-y$ direction.)\label{fig2}}
\end{figure}

Similar to CEP06 Equation 14, for the purposes of our integration of a discretized $p^{i}$ the sphere is divided into spherical shells (analogous to the plane-parallel zones of the original CEP) where $i$ (or $j$) is a shell index. (See Fig. \ref{fig2}.) The integration is broken down into a sum of integrals along different ``cones'' of solid angle (the aforementioned $d\Omega$'s). Note that although $d\Omega$ may seem somewhat conceptually analogous to $d\mu$ in the plane-parallel case, we cannot use the convenient exponential integrals as in the plane-parallel situation, but instead must integrate along each $d\Omega$ separately.

\begin{eqnarray}
\label{partlyDiscreteEqn}
p^{i} = 1 - \int_{4\pi} \left [ \frac{1}{2 \tau^{i,i-1}(\Omega) S^{i}} \sum_{j=1}^{z} S^{j}
\int_{\tau^{i-1}(\Omega)}^{\tau^{i}(\Omega)} d\tau \int_{\tau^{j-1}(\Omega)}^{\tau^{j}(\Omega)} dt \right.\nonumber \\
\left.  \times  \int_{-\infty}^{\infty} \Phi^2(x) dx \left( e^{-|\tau(\Omega)-t(\Omega)|\Phi(x)} \right) \right ]  d\Omega \qquad  \nonumber \\
\end{eqnarray}

Here we have dropped the coordinate subscripts $r,\theta,\phi$ for clarity/simplicity and use shell indices instead (where some shell $i$ contains the point defined by $r,\theta,\phi$).
Note that the ``z'' in the summation, the maximum number of shells along a given $d\Omega$, will be different for each $d\Omega$.
For each ``cone'' of solid angle we will sum $S^{j}(\Omega) e^{-|\tau-t|}$ from ``z'' (the outermost shell) to $i+1$, where $i+1$ is the shell adjacent to shell $i$.

\subsection{Further steps towards implementation}

In actual practice (i.e. computer implementation), the integration of the source function of $d\Omega$ over $4\pi$ steradians will also be done by a discrete summation.

Along any particular element of $d\Omega$ originating in shell $i$, this sum will be: 
\begin{eqnarray}
\sum_{j=1}^{z} S^{j}(d\Omega) ( 1 - e^{-\tau^{j,j-1}(d\Omega)} ) e^{-\tau^{j-1,i}(d\Omega)} \qquad \qquad \nonumber \\
 = \sum_{j=1}^{z} S^{j}(d\Omega) ( 1 - e^{-\tau^{j,j-1}(d\Omega)} ) \prod_{j'=j-1}^{i} e^{-\tau^{j'}(d\Omega)} \quad
\end{eqnarray}

Note that, unlike the plane-parallel case, each $d\tau$ must be calculated explicitly from the geometry and local molecular energy level populations for each shell along the ``pencil'': 
\begin{equation}
d\tau^{j}(d\Omega) = \alpha^{j}(d\Omega) \times ds^{j}(d\Omega),
\end{equation}
where $\alpha^{j}(d\Omega)$ is the absorption coefficient in region $j$ and  $ds^{j}(d\Omega)$ is the distance through that region. Note that this distance may vary over the width of a given $d\Omega$ but we can either approximate using the centerline of $d\Omega$ or derive a $\theta'$-dependent expression and calculate a proper integration to get a mean value over the region. (In our implementation we use the former option, for the sake of simplicity.)

 This integration over solid angle is more tedious than in the plane-parallel case (which was able to use exponential integrals over a zone's $d \tau$) and more computationally costly, but straightforward enough to be feasible.

Turning this all into a fully discrete expression for $p^{i}$ (for shell ``$i$'') we get

\begin{eqnarray}
\label{fullyDiscreteEqn}
p^{i} = 1 - \sum_{\omega=1}^{Z} \left [ \frac{1}{2 \tau^{i,i-1}_{\omega} S^{i}_{\omega}} \sum_{j=1}^{z} S^{j}_{\omega}
( 1 - e^{-(\alpha^{j}_{\omega} \times \Delta s^{j,j-1}_{\omega})} )  \right.\nonumber \\
\left. \times  \prod_{j'=j-1}^{i} e^{-(\alpha^{j'}_{\omega} \times \Delta s^{j', j'-1}_{\omega})}  \times \Delta\Omega_{\omega}\right ] \quad
\end{eqnarray}

Where $Z$ is the maximum numbered spherical shell and all the quantities indicated by the subscript $\omega$ are dependent on a particular direction viewed from a given point (or shell $i$, in a spherically symmetric case). 

\subsection{Asymmetric case: Incident Radiation}

Our motivating interest in this study is comets' comae, which are {\it not} spherically symmetric cases. 

\begin{figure}[h]
\centerline{\includegraphics[trim=15cm 17.5cm 15cm 2cm, clip=true,width=0.5\textwidth]{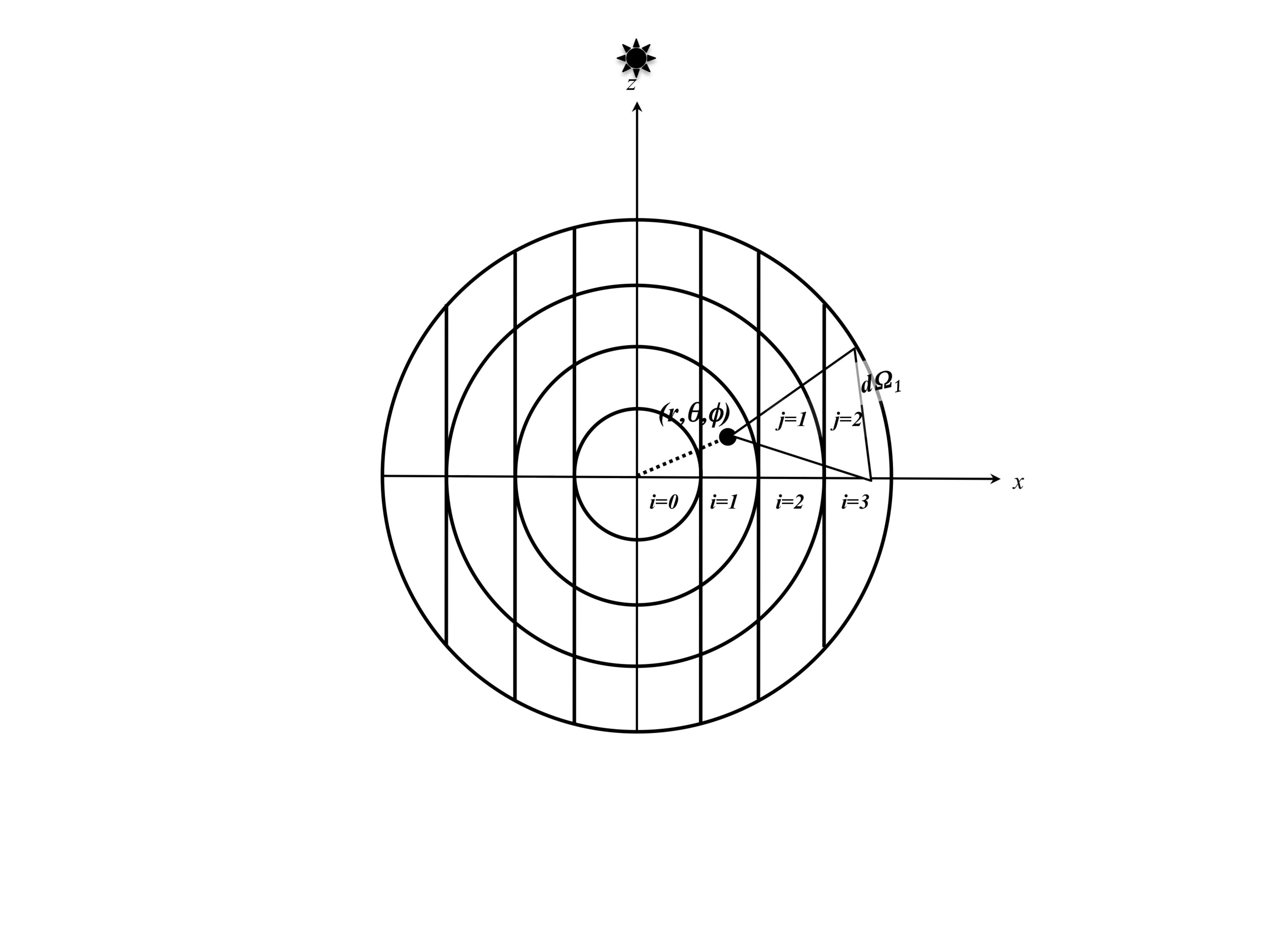}}
\caption{2-D cross section of sphere on $y=0$ plane, viewed from the $-y$ direction, showing division into 4 shells, {\it with} superimposed cylinders along solar ($\hat{z}$) direction.\label{fig3}}
\end{figure}

\begin{figure}[h]
\centerline{\includegraphics[trim=12.5cm 30cm 12.5cm 20cm, clip=true, width=0.5\textwidth]{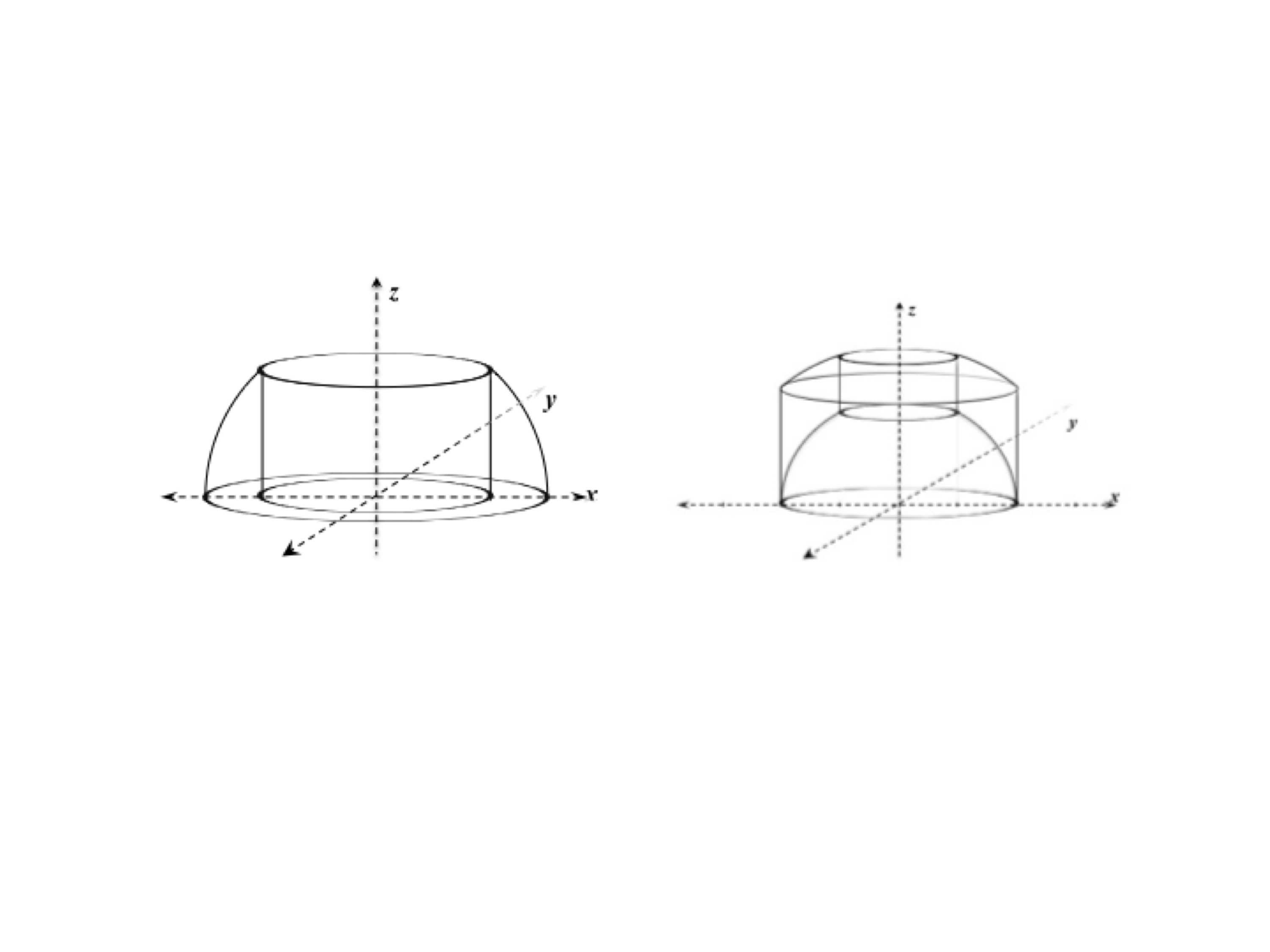}}
\caption{Two examples of different possible shapes of 3D annuli formed by intersecting spheres and cylinders.\label{fig4}}
\end{figure}

To adapt CEP to this asymmetry we divide each shell further into ``regions.''
In the case of comets, one source of asymmetry is incident solar radiation coming from one direction (outside the outermost shell). 
Therefore, the natural way to further divide shells into regions is along lines parallel to the direction of solar radiation (i.e. the center-of-comet-to-center-of-Sun line, which we will arbitrarily label as the $z$-axis.) 
Thus we superimpose a set of co-axial cylinders (with radii equal to corresponding shell radii, for the sake of simplicity) on the shells to divide the coma into regions bounded by two cylinders and two spherical shells. (Note that some regions, specifically those along the $z=0$ plane perpendicular to the solar radiation, are only bounded by an inner cylinder and outer sphere. See Figs. \ref{fig3} \& \ref{fig4}.)

These regions form annuli or rings of unusual, but easily envisioned, cross-sections. (See Fig. \ref{fig4}.)

Incident solar radiation is parallel to the $z$-axis (due to our choice of the $z$ direction). Hence each ray of sunlight travels along the axial direction of a specific cylinder. For each region, the solar radiation absorbed is calculated based on the relevant optical depths along that direction (the $d\tau$'s) of those regions in the same cylinder that are closer to the Sun than the given region. This is similar to the case of external radiation described in CEP06 Appendix A, equations A3 and A4, in which we simply set $\mu_{0} = 1$, due to the above constraint of cylinders being co-axial with the incident solar radiation:

\begin{eqnarray}
\label{externalRadn}
& \bar{J_{e}^{i}} = J_{e} \frac{1}{\tau^{i,i-1}} \left [ \gamma( \tau^{i} ) -  \gamma( \tau^{i-1} ) \right ] \\
 \nonumber \\
&  where  \nonumber \\
 \nonumber \\
& \gamma( \tau ) =  \int_{-\infty}^{\infty} [ 1 - e^{-\tau\Phi(x)} ]  dx  \nonumber
\end{eqnarray}
where $\bar{J_{e}^{i}}$ is the average over a region $i$ of $J_{e}^{i}$, the contribution of external radiation,  $J_{e}$, to the mean intensity of the region, which is to be included in the equations of statistical equilibrium (Equation \ref{bigEqn}) by addition to the ``B'' term.

From a purely radiative standpoint, assuming that within each region there exists uniform density, temperature and other physical conditions, the radiative excitation of molecules (hence, the emission and absorption) in each region/annulus should be equal throughout the region. 
In our expanded CEP implementation, these regions are analogous to zones in the plane-parallel CEP. Each region's radiative effect or contribution to each other region, i.e. the net radiative bracket, must be calculated.
Note that self-irradiation from around an annulus must also be taken into account, as well as irradiation from other regions. Once this calculation is done, the entire region will be  equal with respect to radiative processes (i.e. there is symmetry around the $z$-axis).

\subsection{Further Asymmetry: Coma Morphology}

If models of distantly observed comets were all we needed, this might be sufficient. But we are motivated by the desire to better understand Deep Impact \& EPOXI spectral observations that have very high spatial resolution around the comets' nuclei. (See e.g. Feaga et al. 2007 and Feaga, et al. 2011.)
Therefore the above radiative treatment alone is insufficiently asymmetric to {\it fully} model a cometary coma, when coma morphology is included in the model. The inclusion of morphology undoes the aforementioned symmetry around  the $z$-axis within each annulus/region. These observations are one of the primary motives for this study, and therefore these morphological asymmetries must also be dealt with appropriately in this model.

To model morphological features, we use a cone shape superimposed over the aforementioned divisions into regions. (Other geometric shapes could also have been used. We chose to implement a cone due to its similarity in shape to many observed coma features.) A cone of arbitrary orientation and size with its vertex at the center of the sphere creates intersections with the above-described regions. Each of these is then added as a sub-region, which may have different properties from the surrounding (or subsumed/replaced) region.

Each sub-region can posess different initial conditions from the surrounding region. Thus morphological features, which by their nature tend not to be axisymmetric around our $z$-axis, can be included in the model. It should be noted that these sub-regions are only included as necessary. Thus for those annuli that {\it do} have constant axisymmetric conditions (i.e. no interesting morphological features impinging on them), we can save time and memory computationally by leaving them undivided, as they would have been originally.

\subsection{Implementation: Our Algorithm Described}

Given the above geometric divisions, we have implemented Asymmetric Spherical CEP as follows.
For each region (or sub-region, as applicable) we take representative population values from the coma integration and use these as the basis of an ``initial guess.'' We then make an immediate improvement to the initial guess values by recalculating each region's populations (individually) taking into account the attenuation of incident solar radiation by intervening regions in the solar direction (as per Equation \ref{externalRadn}). These recalculated populations are the values we then use as the initial guess (required by the implementation of Newton's method in Press, et al. 1992) for CEP calculations.
Based on these populations we calculate the necessary source functions, delta-taus, and net radiative brackets, ``$p$,'' as above, for each wavenumber (or line, transition, etc.)
Following the above discretized equation (\ref{fullyDiscreteEqn}) for each region, which in this context we will call the ``recipient'' region, we iterate over all other regions to calculate their contribution to the recipient's $p$. 
Each other region's contribution is essentially its own source function attenuated over the optical depth of all intervening regions along the line of sight between itself and the recipient region, integrated over (or, to simplify, multiplied by) the solid angle subtended by one region from the other. This is then divided by the recipient region's source function and optical depth (along the given line of sight).

To implement this in a practical algorithm of manageable complexity, we make several simplifying approximations.

Due to the $z$-axis symmetry that exists (before adding sub-regions), we can {\it partially} simplify to a two-dimensional diagram in which a region is represented by the cross-section of the annulus in the (arbitrarily chosen) $y=0$ plane. We calculate a region's ``centroid'', i.e. the centroid of its 2-D projected area in this plane which (in our approximation) corresponds to a point  $(r,\theta,\phi)$ in spherical coordinates.

For (cone shaped) sub-regions, which in general do not have their centerline on the X-Z plane, we must use a different centroid. We use the midpoint along the cone's centerline (within region boundaries).

We also choose a series of points distributed evenly along circles parallel to the X-Y plane around each region, which will be the ``starting points'' for calculation of lines of sight (which will terminate at the centroids). These are chosen by rotating a region's centroid around the $z$-axis by multiples of some angle that depends on the size (radius) of the region. The choice of angle is such that the larger a region's size, the more starting points it will have, and thus the region will be divided into more elements of solid angle.

\begin{figure}[h]
\centerline{\includegraphics[trim=20cm 15cm 12.5cm 5cm, clip=true, width=0.5\textwidth]{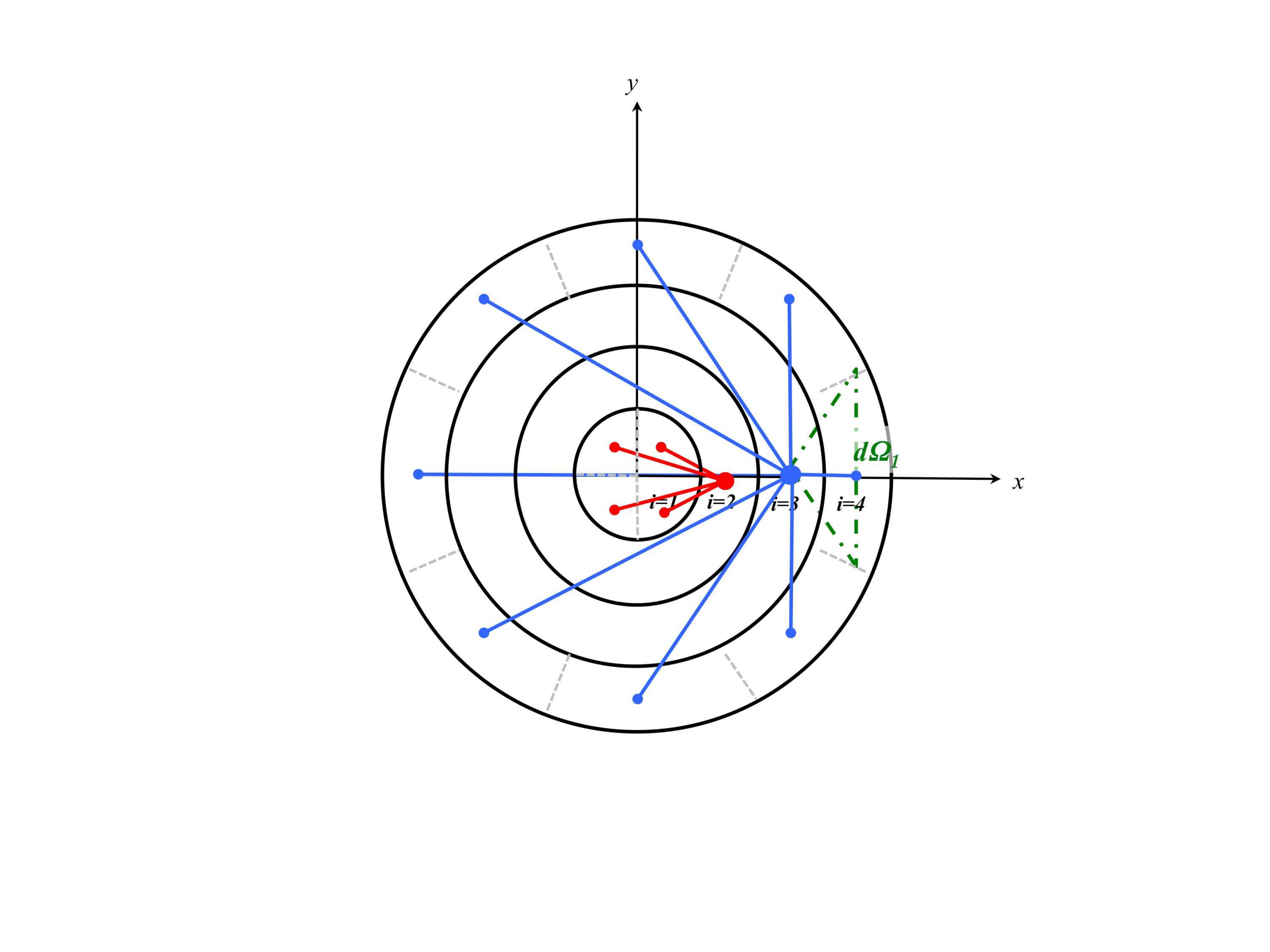}}
\caption{A two-dimensional view from above (i.e. the $+z$ direction) of examples of integration lines in the $X-Y$ plane. Four lines originating in the $i=1$ region and ending at the centroid of the $i=2$ region are shown (in red in the online version), with corresponding division of the $i=1$ circle (shown by dashed grey lines). Eight lines originating in the $i=4$ region and ending at the centroid of the $i=3$ region are shown (in blue in the online version), with corresponding division of the $i=4$ annulus (also shown by dashed grey lines). One example of a corresponding $d \Omega$ is also shown with dotted-dashed lines (in green in the online version). Note that this 2-D diagram only shows horizontal cross-sections of regions, and so regions and shells/annuli are essentially indistingushable in this diagram.\label{fig5}}
\end{figure}

We use the line of sight between the ``centroids'' of regions and this series of starting points to calculate the contributions of every other region (or the region to itself) to a given recipient region's $p$. We calculate the optical depth of each intervening region, along the line of sight, based on the molecular population levels of the intervening regions. (See Fig. \ref{fig5}.) These ``integration lines'' encapsulate the main part (within the square brackets in Equation \ref{fullyDiscreteEqn}) of the calculation of $p$.

\begin{figure}[h]
\centerline{\includegraphics[width=0.5\textwidth]{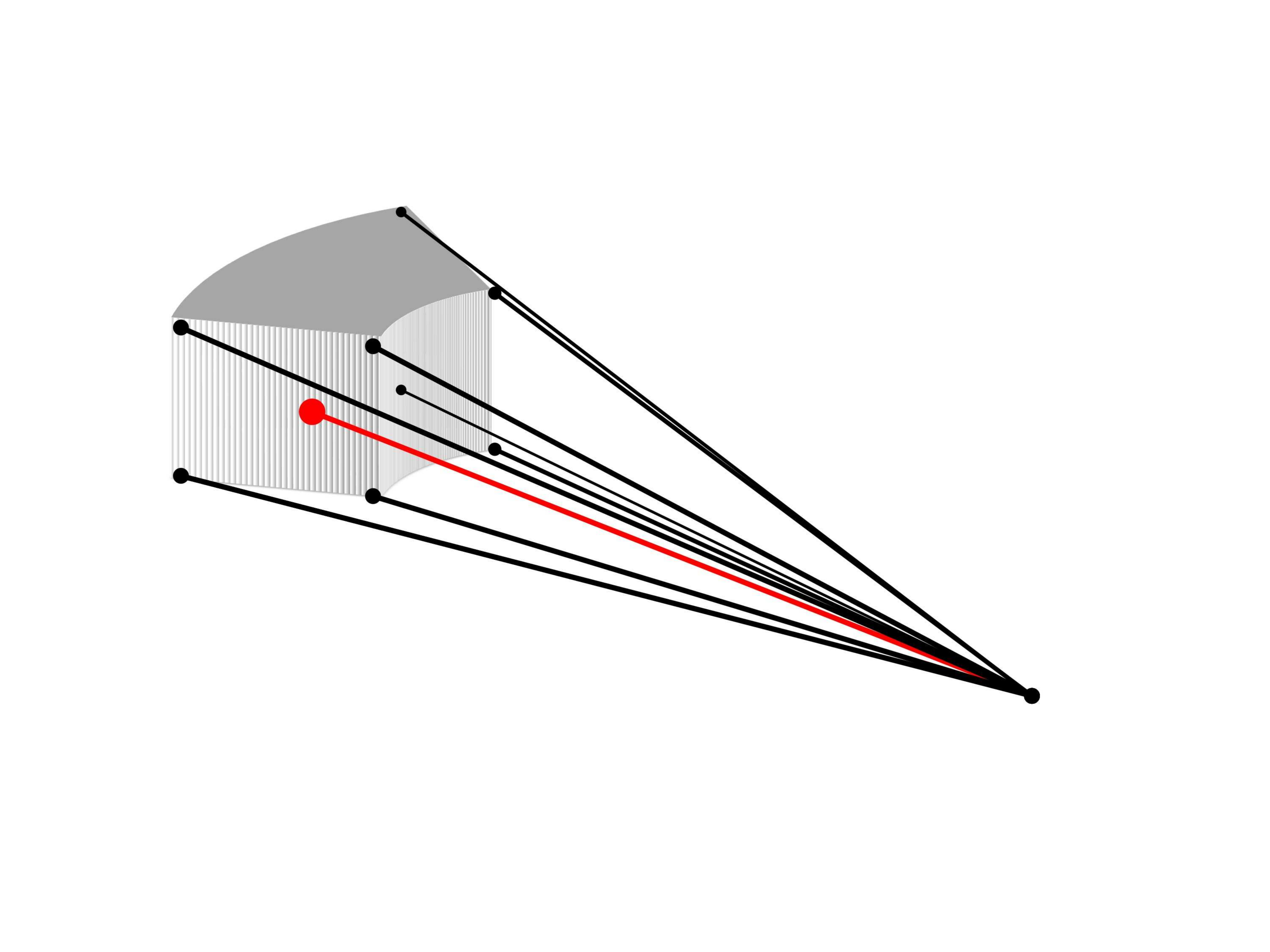}}
\caption{Example illustrating calculation of mean angle. Lines originating from an angular slice of a region's ``corners'' and terminating at another region's centroid are shown in black. The line between the centroid of the ``recipient'' region and a ``start point'' of the other region (the largest dot), corresponds to the relevant integration line (and is shown in red in the online version). The integration line and each of the other eight lines define the angles that are averaged together to get the mean angle $\theta_{mean}$ used to calculate the solid angle subtended by one region as viewed from the other's centroid. Note that not all regions will have eight ``corner points.''\label{fig6}}
\end{figure}

We approximate the solid angle subtended by the integration lines from another region's centroid by integrating $d\Omega = d\phi$ sin$\theta  d\theta$ from zero up to the mean value of the angles between the starting point of that region, the centroid of the recipient region and the multiple ``corners'' of that region around the starting point. (See Fig. \ref{fig6}.) Effectively, this gives a solid angle between regions of $2\pi ( 1 -$ cos$\theta_{mean})$.

Note that due to these approximations, the sum of solid angles over all integration lines between a region and all regions in a given shell exterior to the region is {\it not} necessarily constrained to exactly equal $4\pi$, as it should be in reality. Therefore, in calculating a ``$p$'' value, we sum the solid angles involved and average over that sum instead of $4\pi$ steradians (as Eqns. \ref{analyticalEqn} \& \ref{partlyDiscreteEqn} dictate we should do).

In the limit of arbitrarily small (and numerous) regions, these approximations {\it would} approach a physical situation of arbitrarily precise accuracy. Thus we maintain the ``exactness'' of the CEP method.

\begin{figure}[h]
\centerline{\includegraphics[trim=20.5cm 9cm 0cm 2cm, clip=true, width=0.5\textwidth]{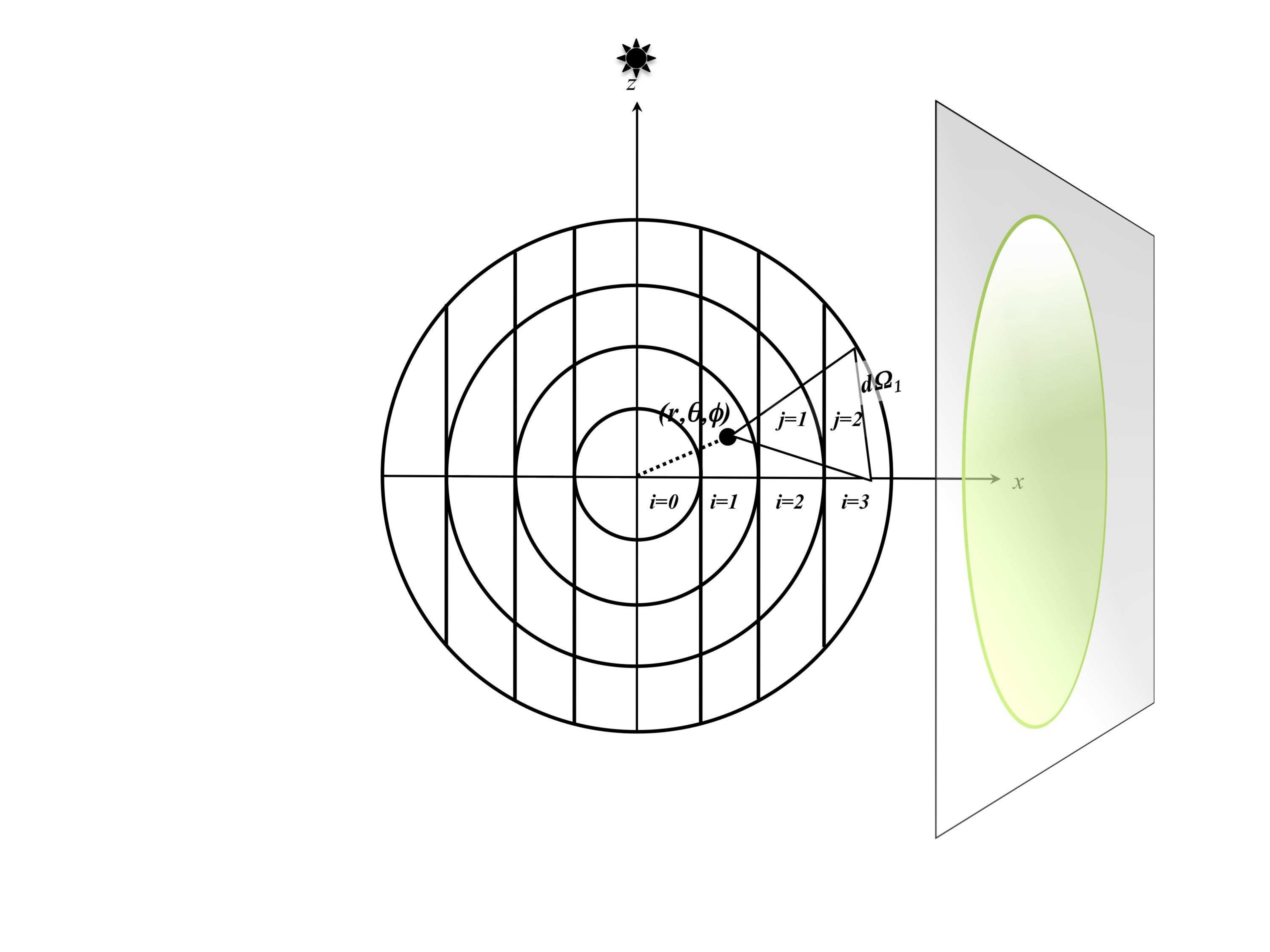}}
\caption{As per Fig. \ref{fig3} above, with observer plane also shown, aligned perpendicular to $X$-axis. (A color image is included in the online version.) \label{fig7}}
\end{figure}

Unlike the plane-parallel situation, the flux exiting the surface of the coma (or other sphere of interest) is not simply a single value (per wavenumber) that has been integrated over angle. In the spherical situation, the resultant intensities form a two-dimensional mapping (in a plane perpendicular to the observer's line of sight. See Fig. \ref{fig7}.) 

In our implementation, this plane is specified by rotation angles, $\theta$, $\phi$ and $\psi$ with the comet's center at the origin, and is assumed to be at a distance $\ge R_{coma}$, the maximum radius of the comet's coma. 
We can also specify the density of and interval between points on this plane for which the output intensities will be calculated.
Each point in this planar mapping shows the intensity (or surface brightness) integrated along a specific line of sight, perpendicular to the plane, through the coma from one side to the other (again, for each wavenumber).

\begin{equation}
I_{surf} = \sum_{i=1}^{z} \left ( S_{i}  \Delta\nu_{i} ( 1 - e^{-\tau^{i}} ) \prod_{j=i-1}^{j=1} e^{-\tau^{j}} \right ) 
\end{equation}

Where $ S_{i}$ is the source function of a region $i$, $\Delta \nu_{i}$ is the line width of wavenumber/frequency $\nu$ in region $i$, and $\tau^{i}$, or $\tau^{j}$, represents the optical depth of wavenumber $\nu$ in region $i$ or $j$ along the relevant line of sight. Indices $i$ and $j$ run from 1 to $z$, where $z$ equals the number of regions along a given line of sight.

Thus the spherical CEP algorithm produces results that could be described as a four-dimensional data ``hypercube'': for each point in the above 2-D mapping, there is a complete (2-D flux vs. wavenumber) spectrum.
 This data can then be presented in multiple formats. Several forms of data presentation for simulating observations are described in the following section.

This is also precisely the output needed to compare with the Deep Impact and EPOXI observations that have been displayed as two-dimensional brightness maps for specific wavelengths or bands.

\section{Some Preliminary Results: Observables for Distant Comets}

We present here examples of model results for three different production rates of carbon monoxide which could be potentially useful for distant (e.g. ground based or orbital telescope) observers of comets. These are modeled using a spherical coma, without any morphological features but including optical depth effects both with respect to incident solar radiation within the coma and with respect to emergent ``observed'' radiation.

The output data from the CEP model can be presented in various ways. Here, we show an example of a band total brightness map (analogous to Feaga, et al. 2007, but for the entire coma), radial profiles of brightness, column density, and g-factors for various azimuthal angles. (These could also be done for individual spectral lines, but in the interests of space and avoiding complexity we have not presented such results here.) We also present spectra integrated over different ``aperture'' sizes. The band total brightness is more likely to be similar to actual observations, but high resolution spectra are possible, even from ground based telescopes (see e.g. DiSanti, et al. 1999 and DiSanti, et al. 2001), in particular for comets close to Earth, which might more closely resemble the latter form of model results.

We also demonstrate the potential usefulness to observers of the ratio of the total brightness of the P branch to that of the R branch of CO to determine whether observations include significant optical depth effects. This may be measurable even with relatively poor spectral resolution.

Many of the model ``input values''  (see Table ~\ref{theoreticalInputParams}) have been chosen so as to facilitate comparisons (of our optically thin cases) with other earlier models. Model parameters that are the same for all the following examples are: Solar distance = 1 AU. Solar flux (over the CO band) is $2.5 \times10^{13}\; photons \, cm^{-2} s^{-1} (cm^{-1})^{-1}$, as per Labs \& Neckel 1968 (and as used by Chin \& Weaver 1984 and Weaver \& Mumma 1984). Gas expansion speed is a constant 0.8 km s$^{-1}$ and the initial gas  temperature at the surface is 200 K. $Q_{H_{2}O} = 10 \times Q_{CO}$, as in Chin \& Weaver 1984. As mentioned above, the radial temperature profile closely follows Combi's 1989 model (see Fig. \ref{combi200}) but scaled to the initial gas temperature at the surface, T$_{surface}$. 

\begin{table*}[]
\centering
\caption{Model input parameters for models of theoretical example comets for CO. }
\begin{tabular}{ll}
\hline
Mean Nucleus Radius & 3 km \\
T$_{surface}$ & 200 K \\
Expansion speed V$_{exp}$ & 0.8 km s$^{-1}$ \\
Q$_{CO}$ & $10^{26}, 10^{27}, 10^{28} s^{-1}$ \\
Band Center Wavenumber & 2149 $cm^{-1}$ \\
Band Center Einstein A & 33 $s^{-1}$ \\
Highest Rot. Level, J$_{max}$ & 20 \\
Solar flux & $2.5 \times10^{13}\; photons \, cm^{-2} s^{-1} (cm^{-1})^{-1}$ \\
$\sigma_{rot}$ & $1.32 \times10^{-14}\;cm^{2}$ \\
\hline
\end{tabular}
\label{theoreticalInputParams}
\end{table*}

The coefficients for CO-H$_2$O collisions (assumed to be the dominant source of collisional excitation) are as per Chin \& Weaver 1984: only rotational excitation and de-excitation are considered. (Vibrational cross sections are about 5 orders of magnitude smaller. See  Weaver \& Mumma, 1984, Table 2.) $C = n_{_{H{2}O}} \, \sigma \, \bar{v}$, where $\bar{v}$ is the average relative speed of the molecules (cm s$^{-1}$), $n_{_{H_{2}O}}$ is the number density of H$_2$O (cm$^{-3}$) and $\sigma$ is the collisional cross section of a given transition of CO (cm$^{2}$). The last value is based on a total cross section of $\sigma_{tot} = 1.32 \times 10^{-14} \; cm^2$, which is apportioned between $\Delta J's$ up to 6 as per Chin \& Weaver's Table 1, which we reproduce here in our Table ~\ref{CWcollisionsTable1}.

\begin{table}[]
\centering
\caption{Reproduction of Chin \& Weaver's Table of CO-H$_{2}$O Collision Cross Section Information.}
\begin{tabular}{ll}
\hline
$\Delta$ J = J$_{upper}$ - J$_{lower}$ &  Fraction of Total De-Excitation \\
\hline
1 ...... & 0.34 \\
2 ...... & 0.25 \\
3 ...... & 0.20 \\
4 ...... & 0.10 \\
5 ...... & 0.07 \\
6 ...... & 0.05 \\
$>$6 ...... & 0 \\
\hline
\end{tabular}
\\Total cross section is always $\sigma_{rot}$ = $1.32 \times10^{-14}\;cm^{2}$.\\ Excitation is derived from de-excitation using \\ detailed balance.\\
\label{CWcollisionsTable1}
\end{table}

\subsection{Brightness Maps}

\begin{figure}[h]
\includegraphics[angle=90,  width=0.5\textwidth]{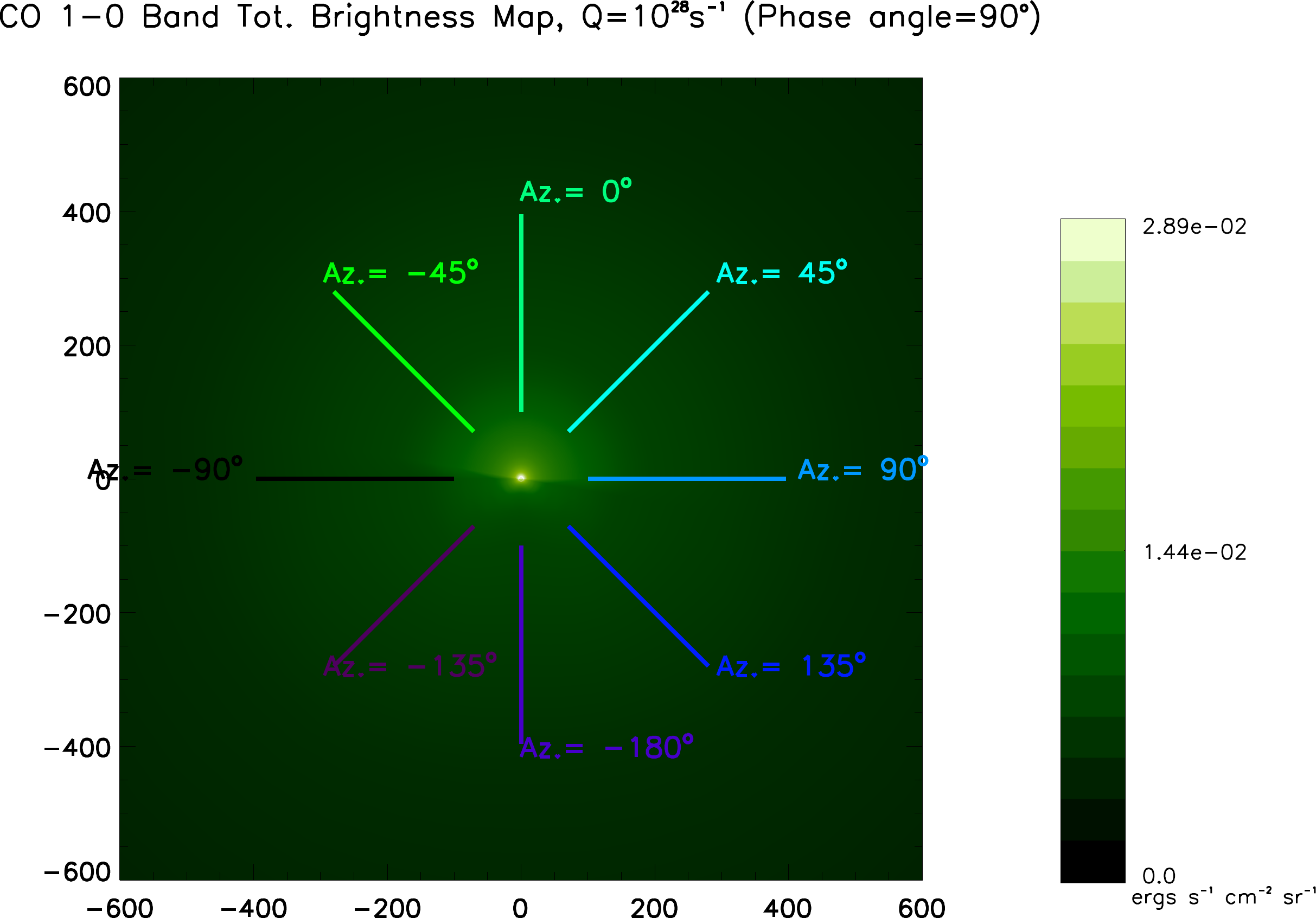} 
\caption{Example of a band total brightness map for the CO 1-0 band, for the inner $\pm$600 km near the nucleus of a theoretical comet with $Q_{CO} = 10^{28} s^{-1}$, viewed from phase angle = 90\degree.  Overlaid radial lines indicate the orientation of azimuthal angles in radial profiles below. The sunward direction is up, i.e. azimuthal angle = 0\degree. Note that within the $\pm$600 km field of view, the brightness never reaches zero (even where it appears to be totally dark). This image is in color in the online version of the article and the colors of the azimuthal angles correspond to those in subsequent radial profiles. \label{brightnessMapQ28} }
\end{figure}

We present here one example of a brightness map of a modeled coma (see Fig. \ref{brightnessMapQ28}). This format of output presentation is most similar to the radiance maps of Feaga, et al. 2007 and A'Hearn, et al. 2011. For a spherical coma with no morphological features, it is rather uninteresting. It is nevertheless included here as a demonstation and used to illustrate the azimuthal angles of the radial profiles presented below with the addition of overlaid lines.
Note that for the $Q_{CO} = 10^{28}$ s$^{-1}$ case, there is some difference in brightness noticeable to the eye between the sunward side (azimuthal angles nearer to zero) and the anti-sunward side, especially in the near-nucleus portion of the image. This is due to the optical depth along the solar direction reducing the excitation and emission from one side of the coma to the other.

\subsection{Radial Profiles: Brightness, Column Density and g-factors}

Abundances of cometary species are frequently calculated from observed fluxes using fluorescence efficiencies, or g-factors. In an optically thin case, the brightness of a given line or band is proportional to the column density of the relevant molecule. In such cases 
$F_{band} = g_{band} \times N$ and $g_{band} = \sum_{band}{ A_{u} \times n_{u} }$ where $F_{band}$ is the band total flux (or brightness), $g_{band}$ the band g-factor, $N$ the total column density, $A_{u}$ the Einstein A coefficient for the relevant transition originating in upper level ``u'' and  $n_{u}$ is the column density of the population of a specific upper level ``u'' (which in our model is numerically approximated as the sum over all regions along a line of sight of the fractional population of level ``u'' times each region's column density).

However, large optical depths will spoil this simple linear relation between column density and brightness. With radiative transfer modeling, it is possible to get a calculated g-factor ($g_{band} = \sum_{band}{ A_{u} \times n_{u} }$) from the model and the ``effective g-factor'', $g_{eff} = F_{band} / N$, which is the actual ratio of brightness to column density. 

\begin{figure}[h]
\centerline{\includegraphics[width=0.5\textwidth]{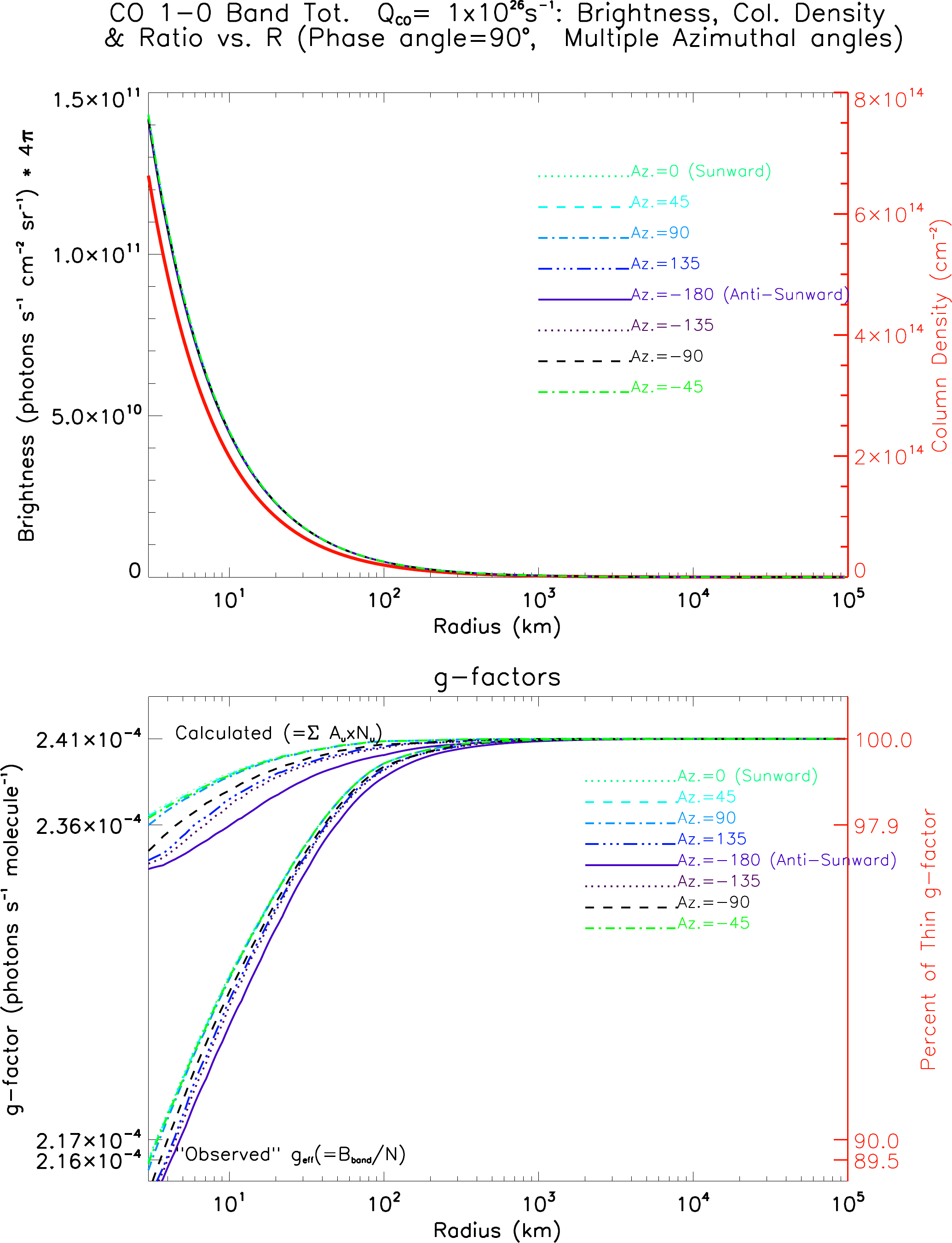}}
\caption{For  $Q_{CO} = 10^{26} s^{-1}$. Upper frame: Radial profile of band total Brightness vs. R (impact paramater) for Phase angle = 90\degree and for multiple Azimuthal angles. Profiles of azimuthal angles (indicated by color coding in the online version) show no variation for this case and overlap, appearing indistinguishable. Column density is included as the bold solid line (red in the online version) using a different y-axis scale. Lower frame: g-factors. Both the calculated g-factor (the higher group of lines) and $g_{eff}$, the observed brightness over column density, plotted with matching styles (colors in the online version) for each azimuthal angle (which also match those in the upper frame). Profiles for 0\degree, $\pm$45\degree, and 90\degree overlap each other almost entirely. \label{fluxVsRphase90multiAzQ26}}
\end{figure}

\begin{figure}[h]
\centerline{\includegraphics[width=0.5\textwidth]{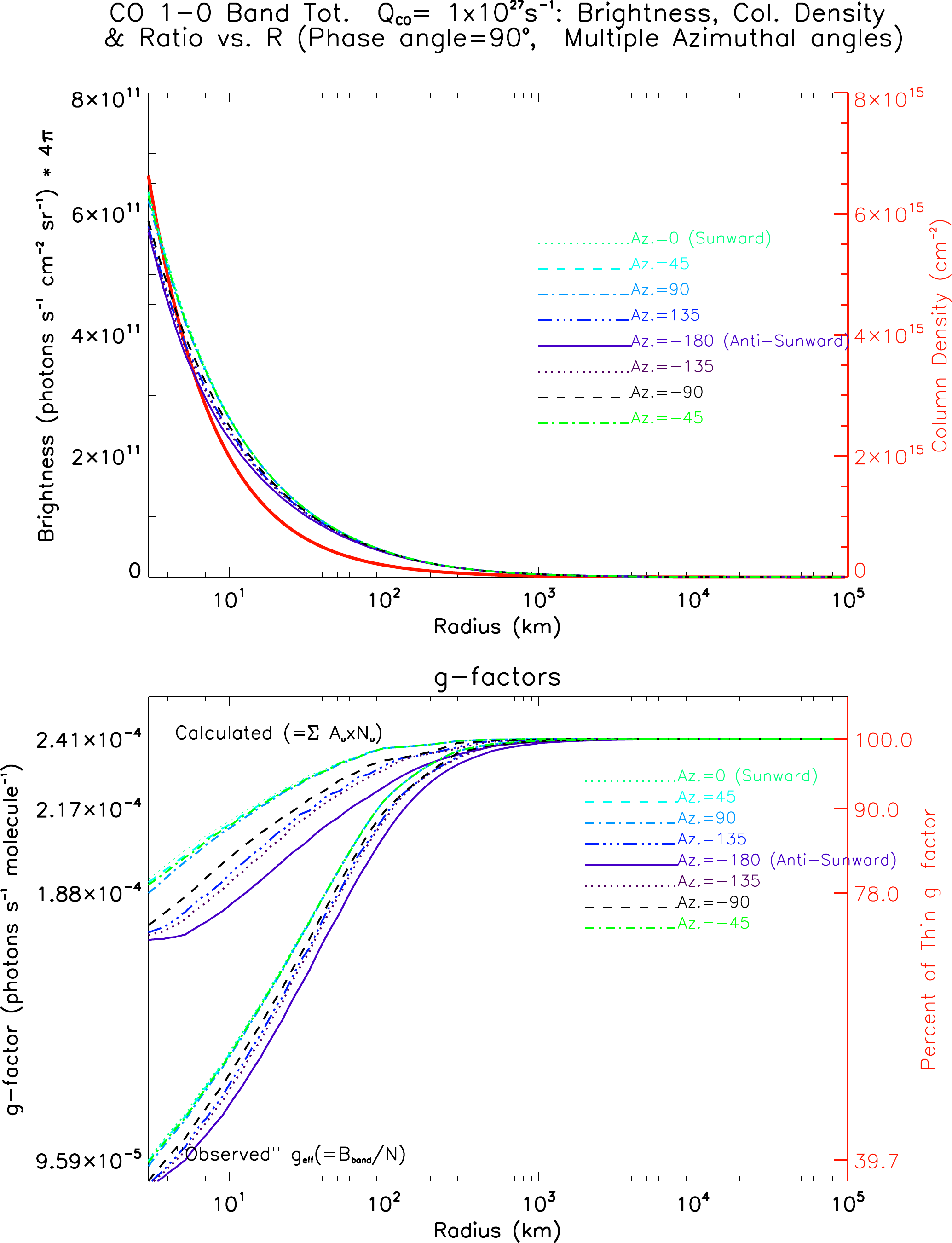}}
\caption{For $Q_{CO} = 10^{27} s^{-1}$. Upper frame: Radial profile of band total Brightness vs. R (impact paramater) for Phase angle = 90\degree and for multiple Azimuthal angles.  Profiles of azimuthal angles (indicated by color coding in the online version) for 0\degree, $\pm$45\degree, and 90\degree overlap each other almost entirely and other brightness profiles are almost indistinguishable. Column density is included as the bold solid line (red in the online version) using a different y-axis scale. Lower frame: g-factors. Both the calculated g-factor (the higher group of lines) and $g_{eff}$, the observed brightness over column density, plotted with matching styles (colors in the online version) for each azimuthal angle (which also match those in the upper frame). Profiles of azimuthal angles 0\degree, $\pm$45\degree, and 90\degree overlap each other almost entirely and are almost indistinguishable. \label{fluxVsRphase90multiAzQ27}}
\end{figure}

\begin{figure}[h]
\centerline{\includegraphics[width=0.5\textwidth]{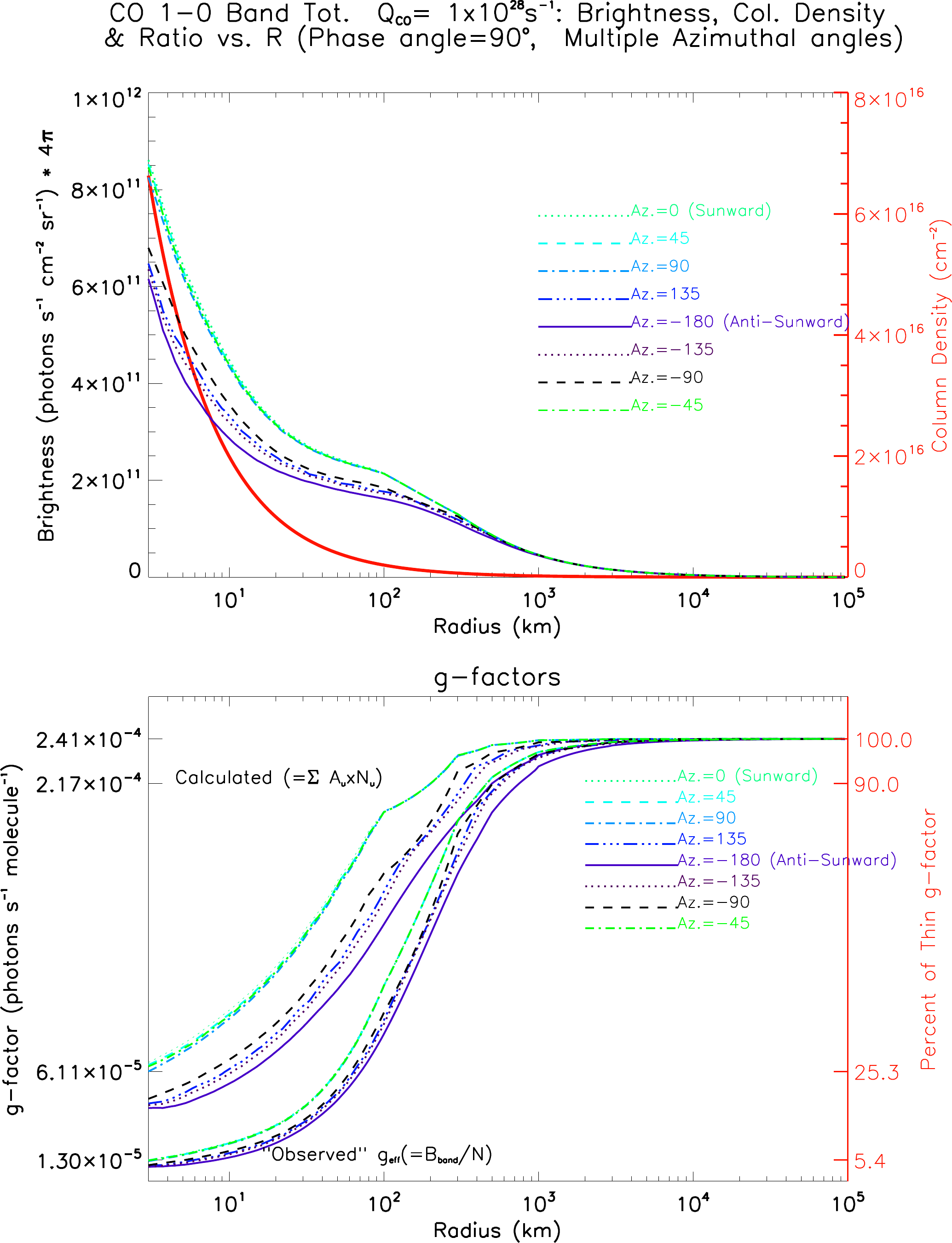}}
\caption{For $Q_{CO} = 10^{28} s^{-1}$. Upper frame: Radial profile of band total Brightness vs. R (impact paramater) for Phase angle = 90\degree and for multiple Azimuthal angles. Profiles of azimuthal angles (indicated by color coding in the online version) for 0\degree, $\pm$45\degree, and 90\degree overlap each other almost entirely. Column density is included as the bold solid line (red in the online version) using a different y-axis scale. Lower frame: g-factors. Both the calculated g-factor (the higher group of lines) and $g_{eff}$, the observed brightness over column density, plotted with matching styles (colors in the online version) for each azimuthal angle (which also match those in the upper frame). \label{fluxVsRphase90multiAzQ28}}
\end{figure}

In Figs. \ref{fluxVsRphase90multiAzQ26}, \ref{fluxVsRphase90multiAzQ27}, and \ref{fluxVsRphase90multiAzQ28} we present all these values together as radial profiles, for theoretical comets of three different production rates, $Q_{CO} = 10^{26},  10^{27}$, and $10^{28} s^{-1}$. (All observed at 1 AU, at a phase angle of 90\degree and multiple azimuthal angles. In all cases $Q_{H_2O} = 10 \times Q_{CO}$.)

In our results, we typically see that $g_{eff}$ does tend towards the calculated g-factor values at larger impact parameters, where optical depth effects are negligible, as would be expected. (The actual numerical values of the ``asymptotic'' band g-factors produced by our model, $2.4 \times 10^{-4} s^{-1}$ per molecule for CO at 1 AU, also agree well with other published values such as those calculated by Chin \& Weaver 1984, Crovisier \& Le Bourlot 1983, and Weaver \& Mumma 1984.)
The actual radii at which this convergence occurs depends primarily on the production rate of a comet. We can use the distance at which $g_{eff} = 0.9 \, g_{thin}$ as a very rough measure of the point where a coma can be considered to transition from optically thick to thin. For the ``thin'' and ``intermediate'' coma models ($Q_{CO} = 10^{26}$ and $Q_{CO} = 10^{27} \, s^{-1}$) the convergence is fairly close to the nucleus, within $\sim$100-200 km. But for the ``thick'' model ($Q_{CO} = 10^{28} s^{-1}$) with its 
high production rate, the ``optically thick regime'' can extend as far as $O( 10^3 )$ km, which can be spatially resolved even in some remote observations. Note that at radial distances very near the nucleus, worrying about optical depth effects may be relevant even for lower production rates.

\subsection{Radial Profiles: Phase and Azimuthal Angular Variations}

Another optical depth effect is variation of brightness (and corresponding g-factor) with varying angles, both phase angle (of observer) and azimuthal angle within any given observation.

The radial profiles in Figs. \ref{fluxVsRphase90multiAzQ26}-\ref{fluxVsRphase90multiAzQ28} demonstrate the azimuthal variation for a single phase (observing) angle. Not surprisingly, the farther a radial profile line is from from the sunward direction, the less bright it is and the lower its g-factors for a given distance from the nucleus. However, the degree to which the actual g-factor varies is of note. Even for the only moderately thick case of $Q_{CO} = 10^{27} s^{-1}$ for radii $\le$ 100 km, there is a difference of as much as $\sim$10\% between sunward and anti-sunward directions. The effect is even more pronounced for the thicker case of $Q_{CO} = 10^{28} s^{-1}$.

\begin{figure}[h]
\centerline{\includegraphics[width=0.5\textwidth]{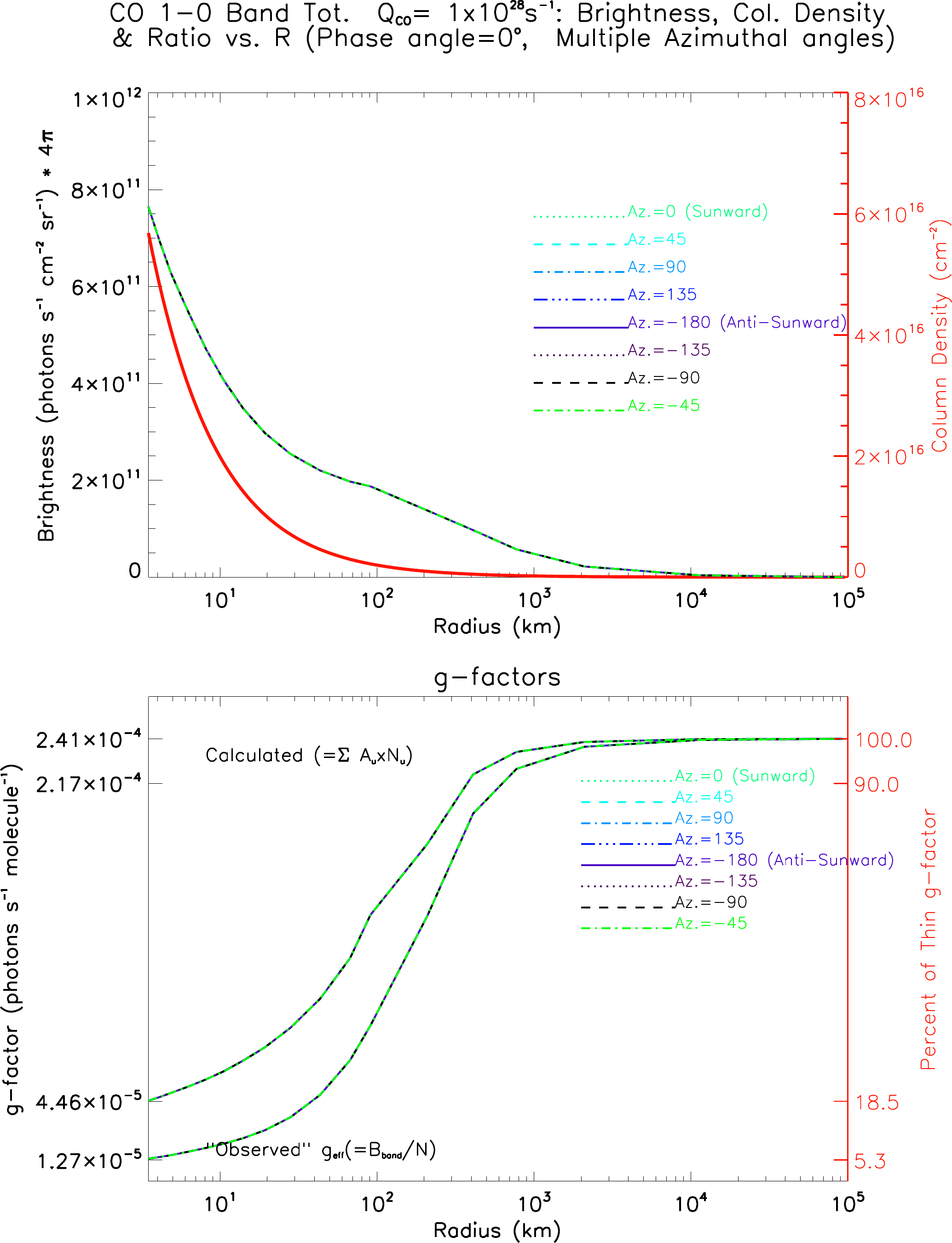} }
\caption{For Phase angle = 0\degree, $Q_{CO} = 10^{28} s^{-1}$. Upper frame: Radial profile of band total Brightness vs. R (impact paramater) for Phase angle = 90\degree for multiple Azimuthal angles. Profiles of azimuthal angles (indicated by color coding in the online version) show no variation for this case (as should be expected) and overlap, appearing indistinguishable. Column density is included as the bold solid line (red in the online version) using a different y-axis scale. Lower frame: g-factors. Both the calculated g-factor (the higher group of lines) and $g_{eff}$, the observed brightness over column density, plotted with matching styles (colors in the online version) for each azimuthal angle (which also match those in the upper frame). \label{fluxVsRphase0multiAzQ28}}
\end{figure}

\begin{figure}[h]
\centerline{\includegraphics[width=0.5\textwidth]{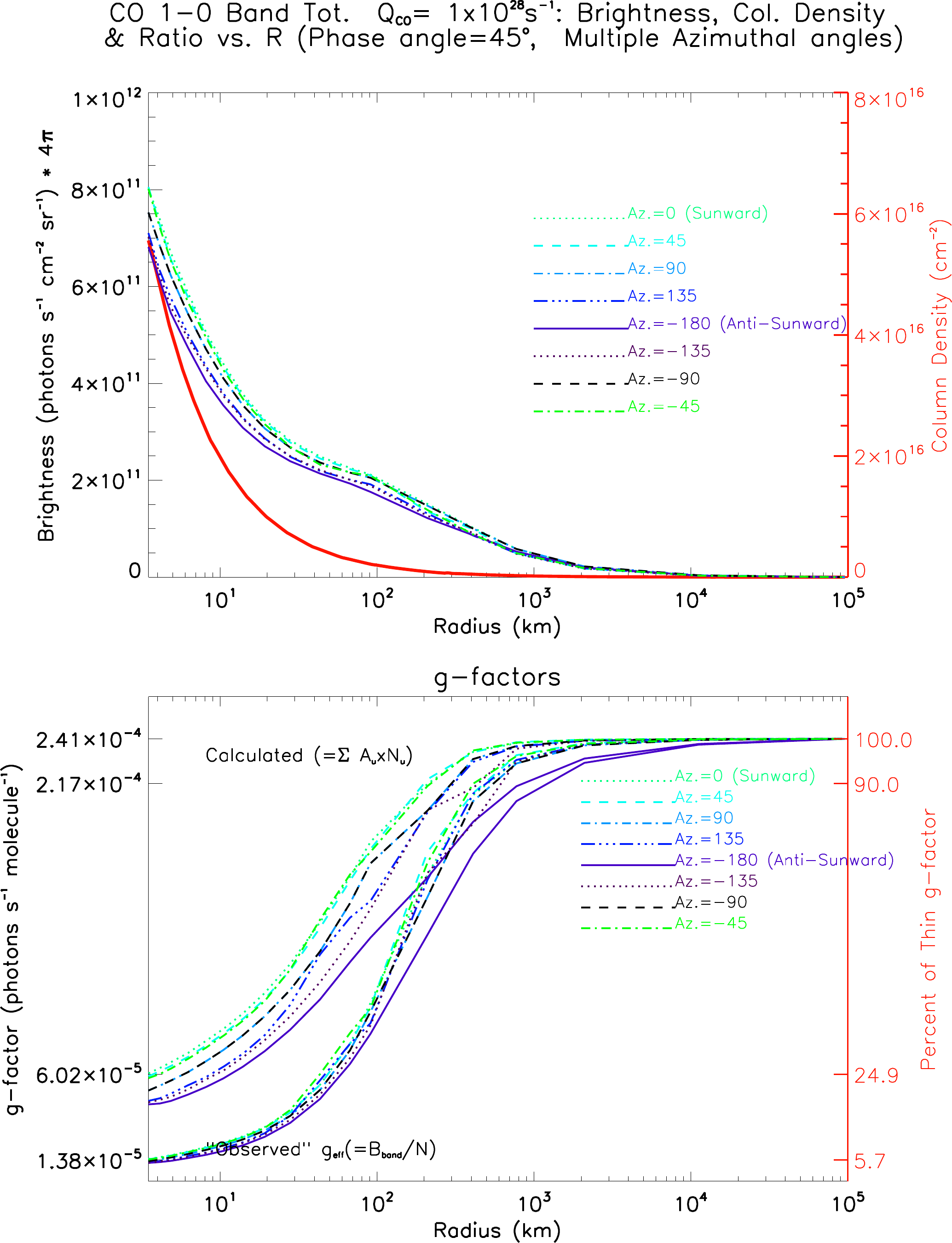}}
\caption{For Phase angle = 45\degree, $Q_{CO} = 10^{28} s^{-1}$. Upper frame: Radial profile of band total Brightness vs. R (impact paramater) for Phase angle = 90\degree for multiple Azimuthal angles. Profiles of azimuthal angles (indicated by color coding in the online version) are slightly displaced for viewing purposes. Column density is included as the bold solid line (red in the online version) using a different y-axis scale. Lower frame: g-factors. Both the calculated g-factor (the higher group of lines) and $g_{eff}$, the observed brightness over column density, plotted with matching styles (colors in the online version) for each azimuthal angle (which also match those in the upper frame). \label{fluxVsRphase45multiAzQ28}}
\end{figure}

\begin{figure}[h]
\centerline{\includegraphics[width=0.5\textwidth]{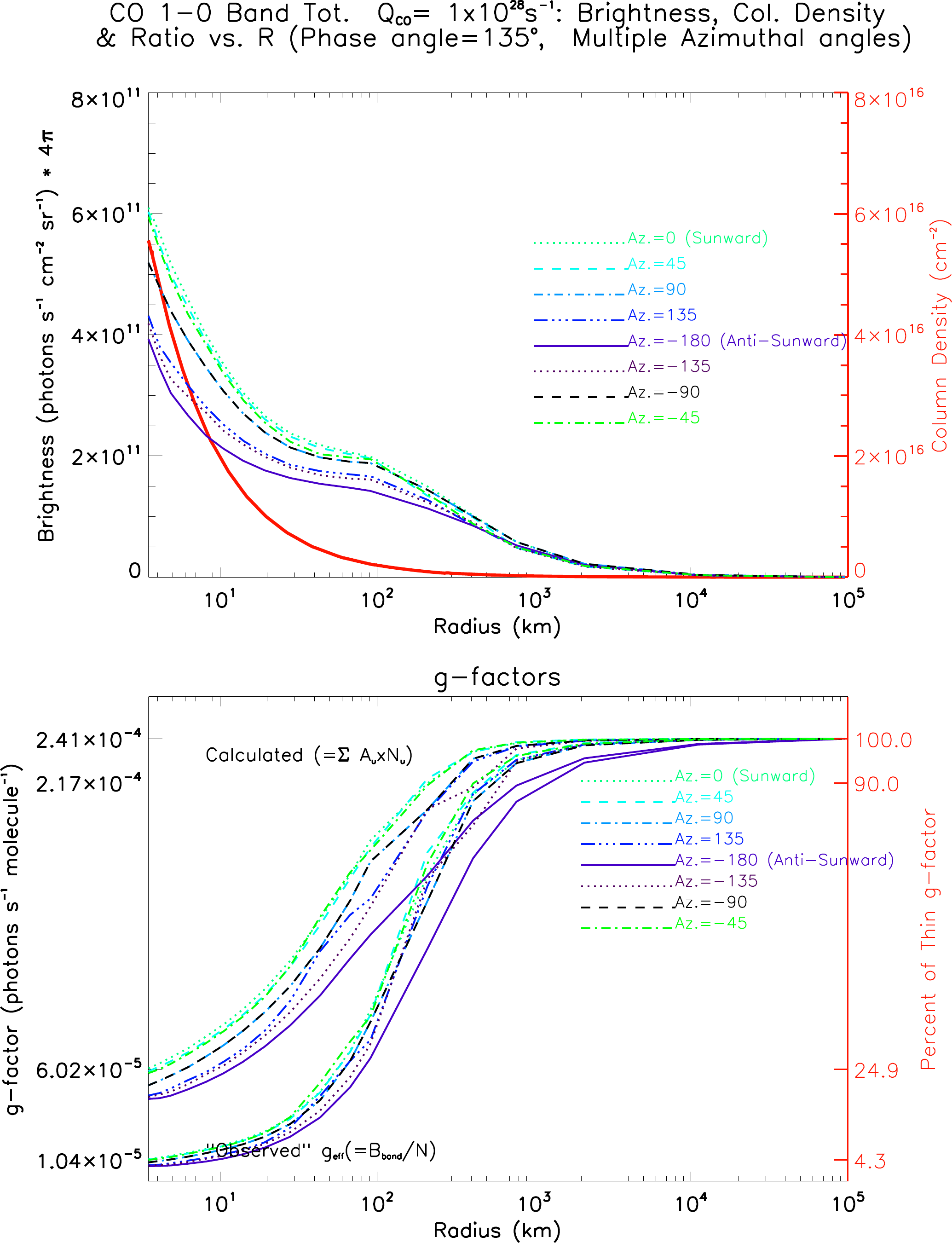}}
\caption{For Phase angle = 135\degree, $Q_{CO} = 10^{28} s^{-1}$. Upper frame: Radial profile of band total Brightness vs. R (impact paramater) for Phase angle = 135\degree for multiple Azimuthal angles. Profiles of azimuthal angles (indicated by color coding in the online version) are slightly displaced for viewing purposes. Column density is included as the bold solid line (red in the online version) using a different y-axis scale. Lower frame: g-factors. Both the calculated g-factor (the higher group of lines) and $g_{eff}$, the observed brightness over column density, plotted with matching styles (colors in the online version) for each azimuthal angle (which also match those in the upper frame). \label{fluxVsRphase135multiAzQ28}}
\end{figure}

\begin{figure}[h]
\centerline{\includegraphics[width=0.5\textwidth]{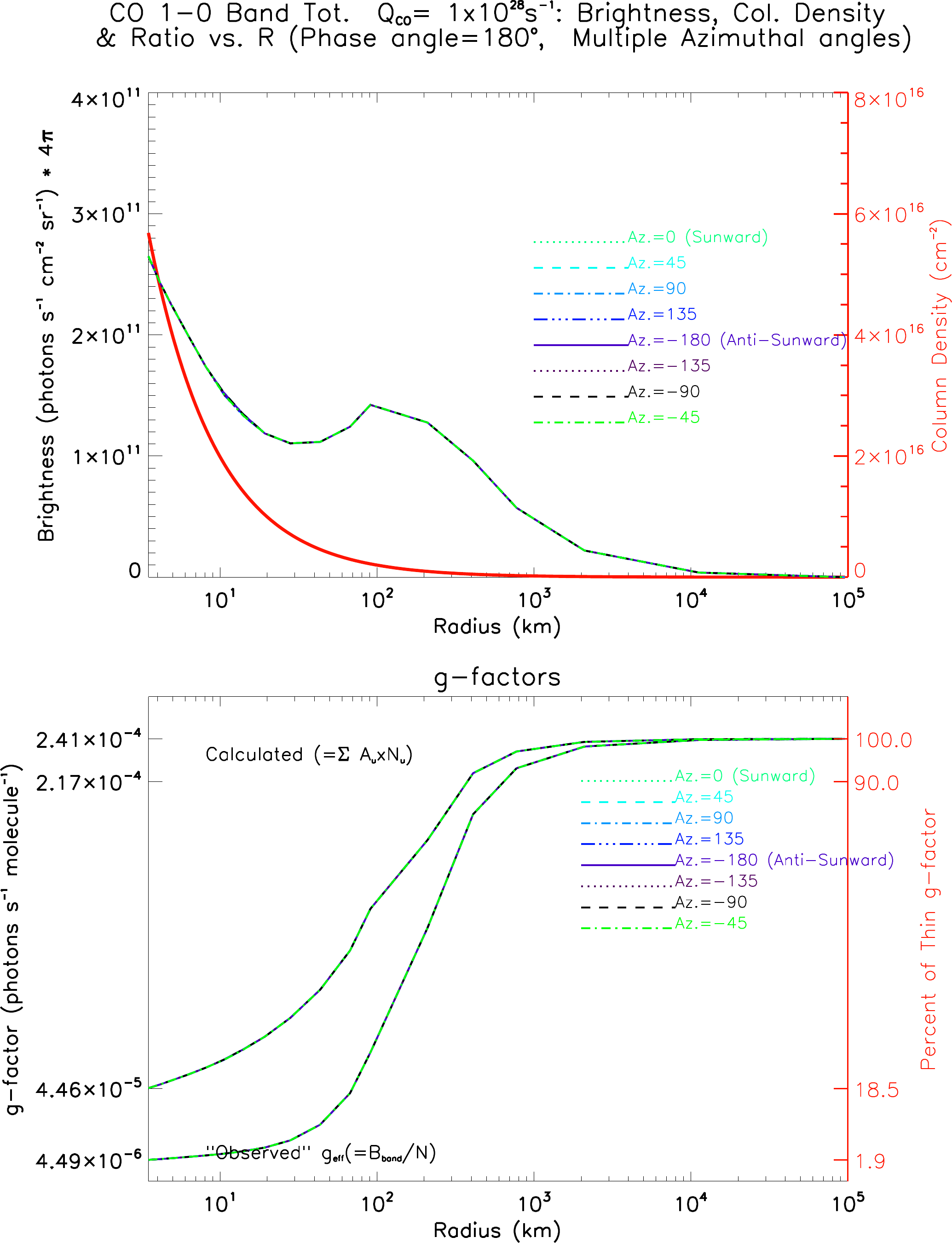}}
\caption{For Phase angle = 180\degree, $Q_{CO} = 10^{28} s^{-1}$. Upper frame: Radial profile of band total Brightness vs. R (impact paramater) for Phase angle = 180\degree for multiple Azimuthal angles. Profiles of azimuthal angles (indicated by color coding in the online version) show no variation for this case (as should be expected) and overlap, appearing indistinguishable. Column density is included as the bold solid line (red in the online version) using a different y-axis scale. Lower frame: g-factors. Both the calculated g-factor (the higher group of lines) and $g_{eff}$, the observed brightness over column density, plotted with matching styles (colors in the online version) for each azimuthal angle (which also match those in the upper frame). \label{fluxVsRphase180multiAzQ28}}
\end{figure}

In Figs. \ref{fluxVsRphase90multiAzQ28}-\ref{fluxVsRphase180multiAzQ28} we present profiles of brightness (and column density) for model results observed from different phase angles, ranging from 0\degree to 180\degree, for the optically thicker case of $Q_{CO} = 10^{28} s^{-1}$.
Each plot includes multiple azimuthal angles, which would all be visible simultaneously in a wide field observation (i.e. including the entire coma) from each given phase angle. (The above plot for 90\degree phase angle for $Q_{CO} = 10^{28} s^{-1}$, Fig. \ref{fluxVsRphase90multiAzQ28}, should be considered part of this series as well.)

Observations with a slit spectrometer with sufficient spatial resolution (see, e.g. DiSanti, et al. 1999 ) might observe along one specific azimuthal angle and thus see possible variations along the slit with sufficiently high spatial resolution.

The most obvious effect seen at a glance in these figures is the spread among azimuthal angles for a given phase angle. As would be expected, the 0\degree and 180\degree phase angles (sunward and anti-sunward) have no real azimuthal variation. 
From phase angle 45\degree to 90\degree to 135\degree there is a progression: the azimuthal lines get spread out farther from each other, as well as noticably dropping in brightness for those lines farther from the sunward side.

With respect to total brightness, the phase angles 0\degree, 45\degree and 90\degree, produce roughly equal peak brightness for their {\it strongest} azimuthal profiles (the more sunward directions, at the nucleus grazing radius) of about $\sim 8 \times 10^{11}$ photons s$^{-1}$ cm$^{2}$ sr$^{-1}$ For the 135\degree phase angle, the peak values are about  $\sim 6 \times 10^{11}$. Most significantly, for the 180\degree phase angle, the peak values are about  $\sim 3 \times 10^{11}$ - only about half as bright as at other phase angles.

With respect to ``effective'' (or ``observed'') g-factors, the minimum values in the most optically thick regions (again, at the nucleus-grazing radii) also show a trend from sunward to anti-sunward. For the 0\degree, 45\degree, and 90\degree phase angles, the minimum values of the ratio of the band total flux over the column density are roughly $1.3 \times 10^{-5}$ photons s$^{-1}$ molecule$^{-1}$. From 90\degree, through 135\degree and down to 180\degree the values go down monotonically to a minimum of about $4.5 \times 10^{-6}$. 
This trend is due to a combination of two optical depth effects. The first is attenuation of incident solar light from the sunward to anti-sunward sides of the coma, leading to less fluorescent pumping, and thus less emission, towards the anti-sunward direction. Second, whatever emission there is is more likely to ``escape'' the coma over shorter optical depths - i.e. closer to where it is emitted. Thus the already greater emission of the sunward regions is also more likely to be observed along azimuthal directions closer to sunward.

However, the similarly located values for ``actual'' calculated g-factors do {\it not} follow a similar simple monotonic trend. The greatest values for a given phase angle rise from 0\degree through 45\degree and peak for phase angle 90\degree. From 90\degree, through 135\degree down to 180\degree they fall through the same values, creating a symmetric peak around 90\degree.
For 90\degree (see Fig. \ref{fluxVsRphase90multiAzQ28}) there are two ``clusters'' of lines. The higher one corresponding to more sunward azimuthal angles (between $\pm45$) and the lower one to other angles. The upper cluster's (minimum) values are actually the highest values in this comparison, about $\sim 6.1 \times 10^{-5}$. The lower cluster of values at 90\degree is still greater than either angle 0\degree or 180\degree, which are the lowest for any phase angle, notwithstanding the azimuthal spread for the other phase angles. (The calculated values also show considerably more spread among azimuthal angles than the profiles of flux over column density.)
This symmetrical and non-monotonic pattern is less intuitive than the trend of $g_{eff}$ above. Yet it is clearly understood in light of the fact that these values are based only on the actual population distributions in different regions and do {\it not} include optical depth effects on the emergent radiation. Thus observing from azimuthal angles 0\degree and 180\degree are sampling exactly the same lines of sight and regions' populations, including the darkest (i.e. least excited populations) of the anti-sunward side of the coma. The same is essentially true for 45\degree and 135\degree (due to symmetry around the z-axis) but they do not sample the darkest parts of the anti-sunward directions (and the differences in populations are less between outermost regions on the sunward side among azimuthal angles between $\pm$45 -- they are all experiencing direct solar illumination). For 90\degree the higher cluster is sampling from more excited and higher emmision populations than the lower cluster, and consistently so all the way along their lines of sight (which is not true for 45\degree and 135\degree). Thus the sunward cluster of lines for 90\degree is the brightest seen, and the anti-sunward cluster values falls between values of the 0\degree or 180\degree and the 45\degree or 135\degree lines.

Lastly, the profiles for $Q_{CO} = 10 ^{28} s^{-1}$ have a ``bump'' in brightness in the vicinity of $\sim$ 100 $\sim$ 1000 km, most easily visible for the 180\degree plot, but also present for the other phase angles (and growing in size from 0\degree up to 180\degree). This is mostly due to the temperature profile reaching its minimal values at these radii in conjunction with the higher density of the $Q_{CO} = 10 ^{28} s^{-1}$ case. The higher density leads to this still being a collsionally dominated regime, and the low temperatures lead to the lowest population levels being most highly populated. These levels also have the highest Einstein A values, thus leading to higher overall number of photons emitted for the same number of molecules. The effect is greatest for the 180\degree view due to a cumulative effect -- the lines of sight all sample the most dense and cold regions at these radii. The same extreme effect is not seen for the 0\degree phase angle due to the overall greater fluorescent excitation of the sunward regions dominating it. (Note that the difference in total brightness between the two azimuthal angles at these radii is about a factor of two.)

\subsection{Aperture Averaged Spectra}

If one is observing with high spectral resolution but low spatial resolution, the spectra observed will be the sum of as much of the coma as fills the field of view. To model this, we have simulated ``aperture averaged'' spectra, where the ``aperture'' controls the area of the coma sampled. Our apertures are square boxes and are all centered exactly on the center of the comet, and sample a nucleus centered area equal to the square of the ``aperture size'' over which we average the brightness. We present a series of apertures from $2\times10^1$ km (very near the nucleus) through $2\times10^5$ km (the whole coma) for each of the three production rates. (All these example spectra are modeled at a heliocentric distance of 1 AU and phase angle 90\degree. $Q_{H_{2}O} = 10 \times Q_{CO}$, as in Chin \& Weaver 1984.)

\begin{figure*}[!h]
\centering
\subfigure[For $Q_{CO} = 10^{26}$ s$^{-1}$. Aperture = 20km.]{\label{aperture-R10-Q26}
          \includegraphics[width=0.35\textwidth,height=0.45\textwidth,angle = 90]{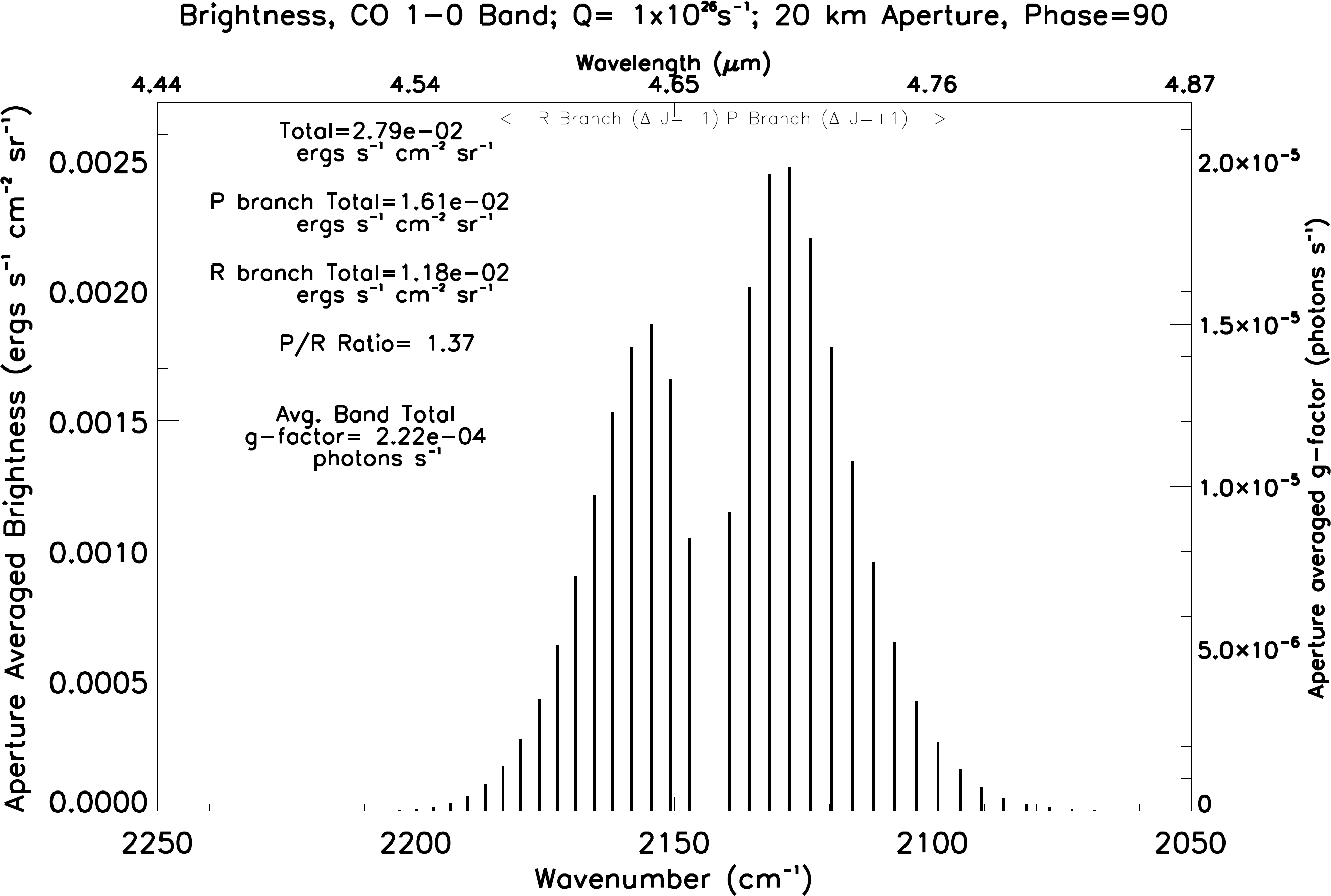}}
	\vspace{0.0in}
\subfigure[For $Q_{CO} = 10^{26}$ s$^{-1}$. Aperture = 100km.]{\label{aperture-R50-Q26}
          \includegraphics[width=0.35\textwidth,height=0.45\textwidth,angle = 90]{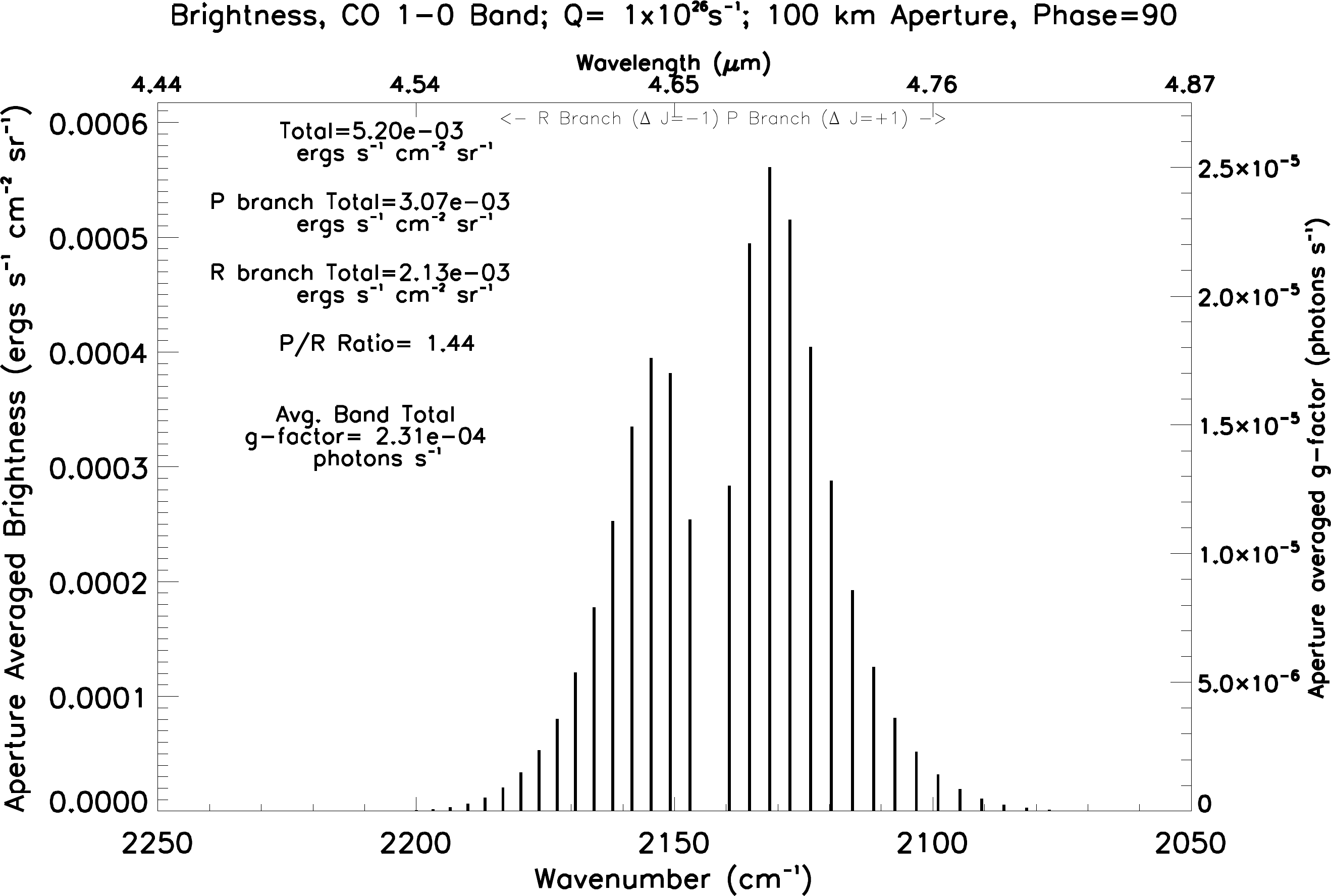}}
	\vspace{0.0in}
\subfigure[For $Q_{CO} = 10^{26}$ s$^{-1}$. Aperture = 200km.]{\label{aperture-R100-Q26}
          \includegraphics[width=0.35\textwidth,height=0.45\textwidth,angle = 90]{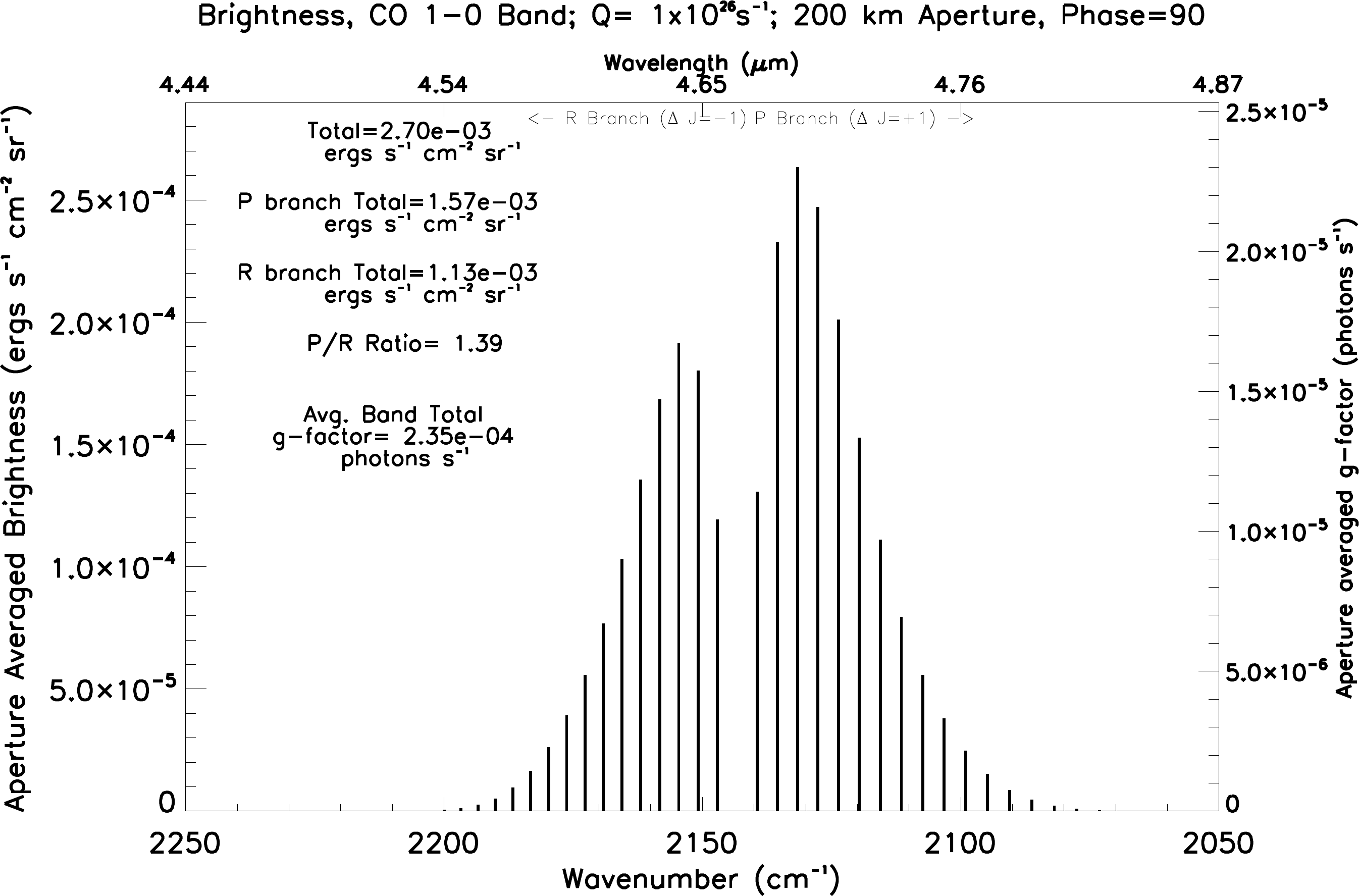}}
	\vspace{0.0in}
\subfigure[For $Q_{CO} = 10^{26}$ s$^{-1}$. Aperture = 2,000km.]{\label{aperture-R1000-Q26}
          \includegraphics[width=0.35\textwidth,height=0.45\textwidth,angle = 90]{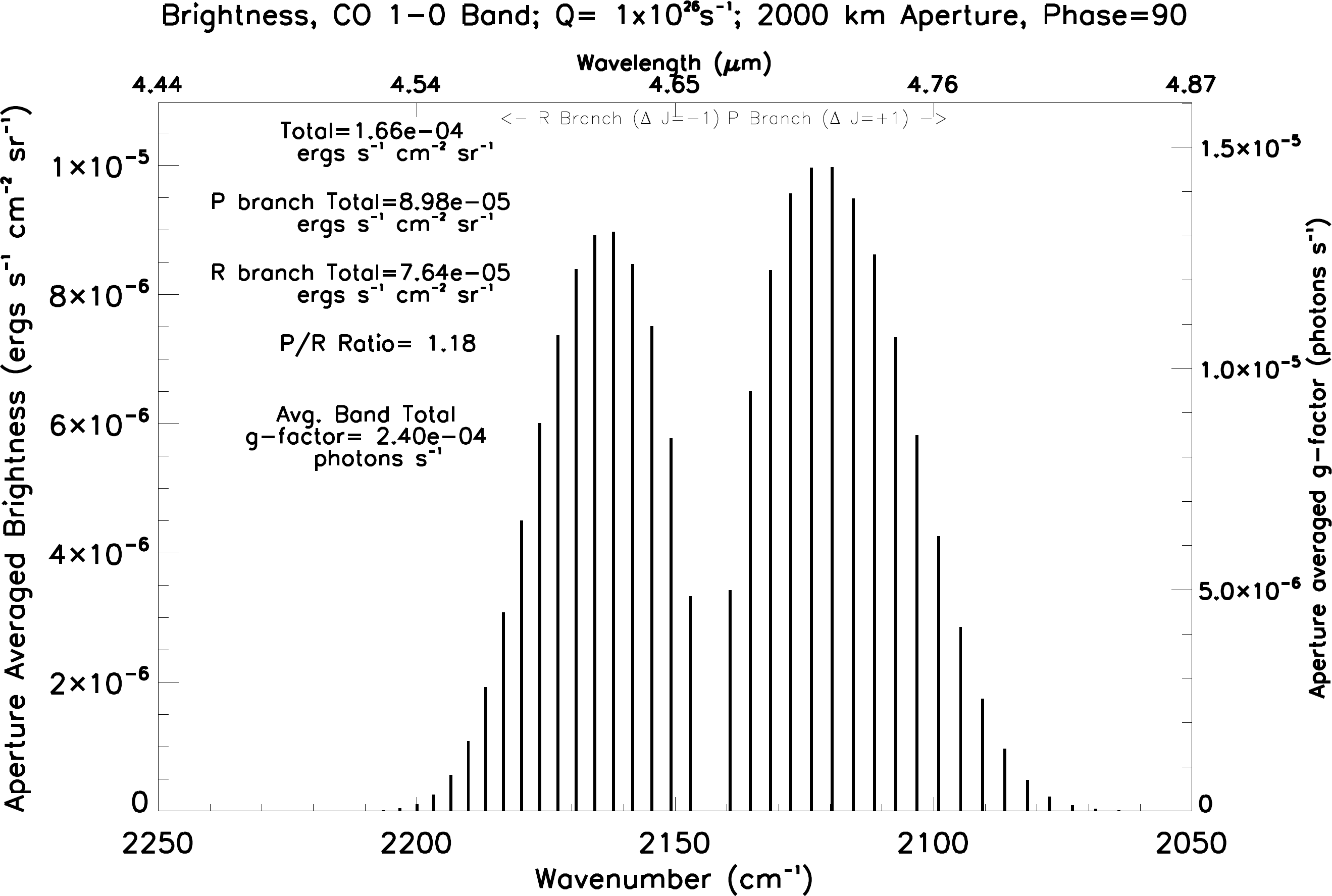}}
	\vspace{0.0in}
\subfigure[For $Q_{CO} = 10^{26}$ s$^{-1}$. Aperture = 20,000km.]{\label{aperture-R10000-Q26}
          \includegraphics[width=0.35\textwidth,height=0.45\textwidth,angle = 90]{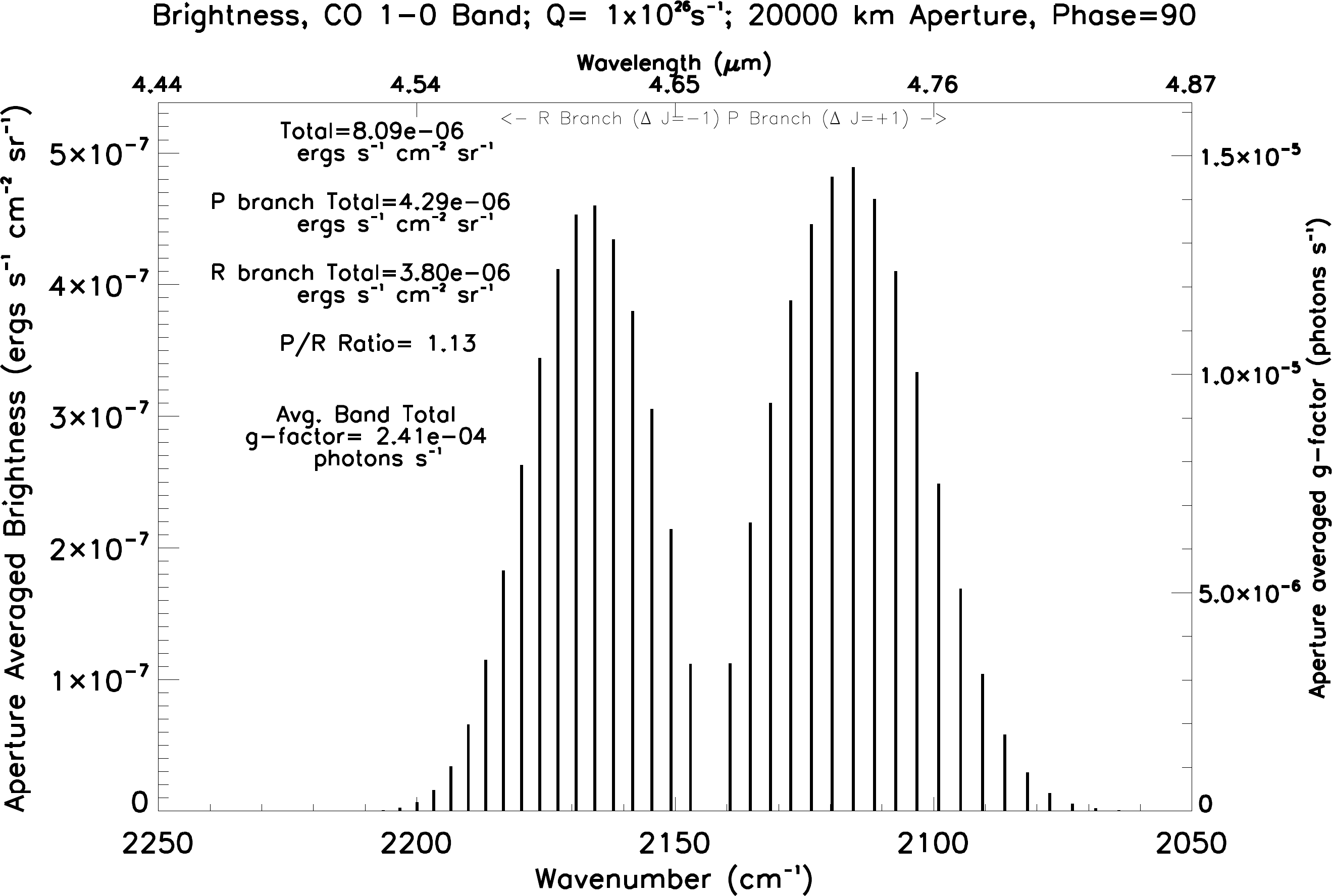}}
	\vspace{0.0in}
\subfigure[For $Q_{CO} = 10^{26}$ s$^{-1}$. Aperture = 200,000km.]{\label{aperture-R100000-Q26}
          \includegraphics[width=0.35\textwidth,height=0.45\textwidth,angle = 90]{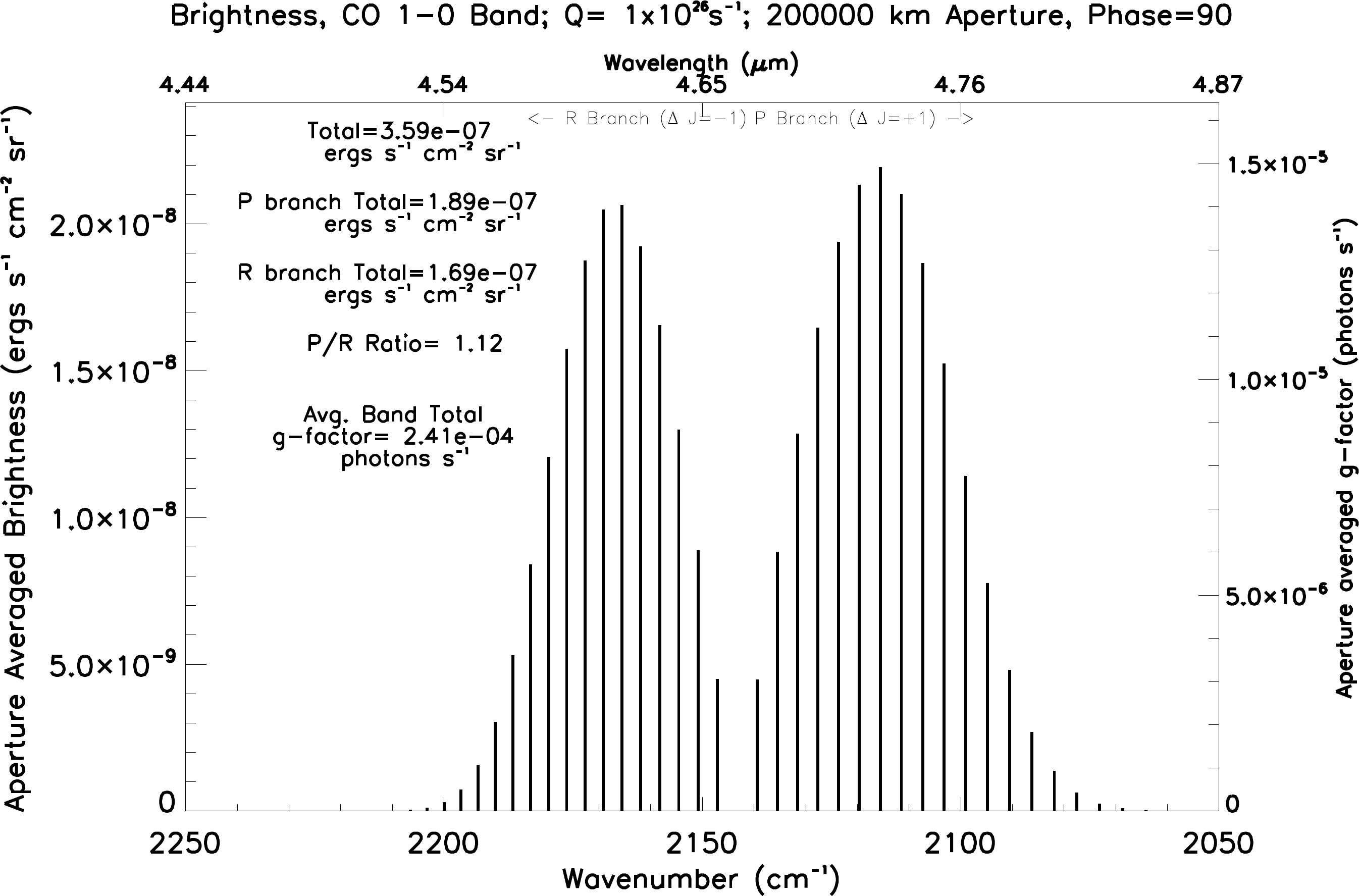}}
	\vspace{0.0in}
\caption{Aperture integrated spectra for $Q_{CO} = 10^{26}$ s$^{-1}$. Left side y-axis is aperture averaged brightness. Right y-axis is effective (line) g-factor (brightness/column density). Totals are indicated on each graph.\label{Q26spectra}}
\end{figure*}

\begin{figure*}[!ht]
\centering
\subfigure[For $Q_{CO} = 10^{27}$ s$^{-1}$. Aperture = 20km.]{\label{aperture-R10-Q27}
          \includegraphics[width=0.35\textwidth,height=0.45\textwidth,angle = 90]{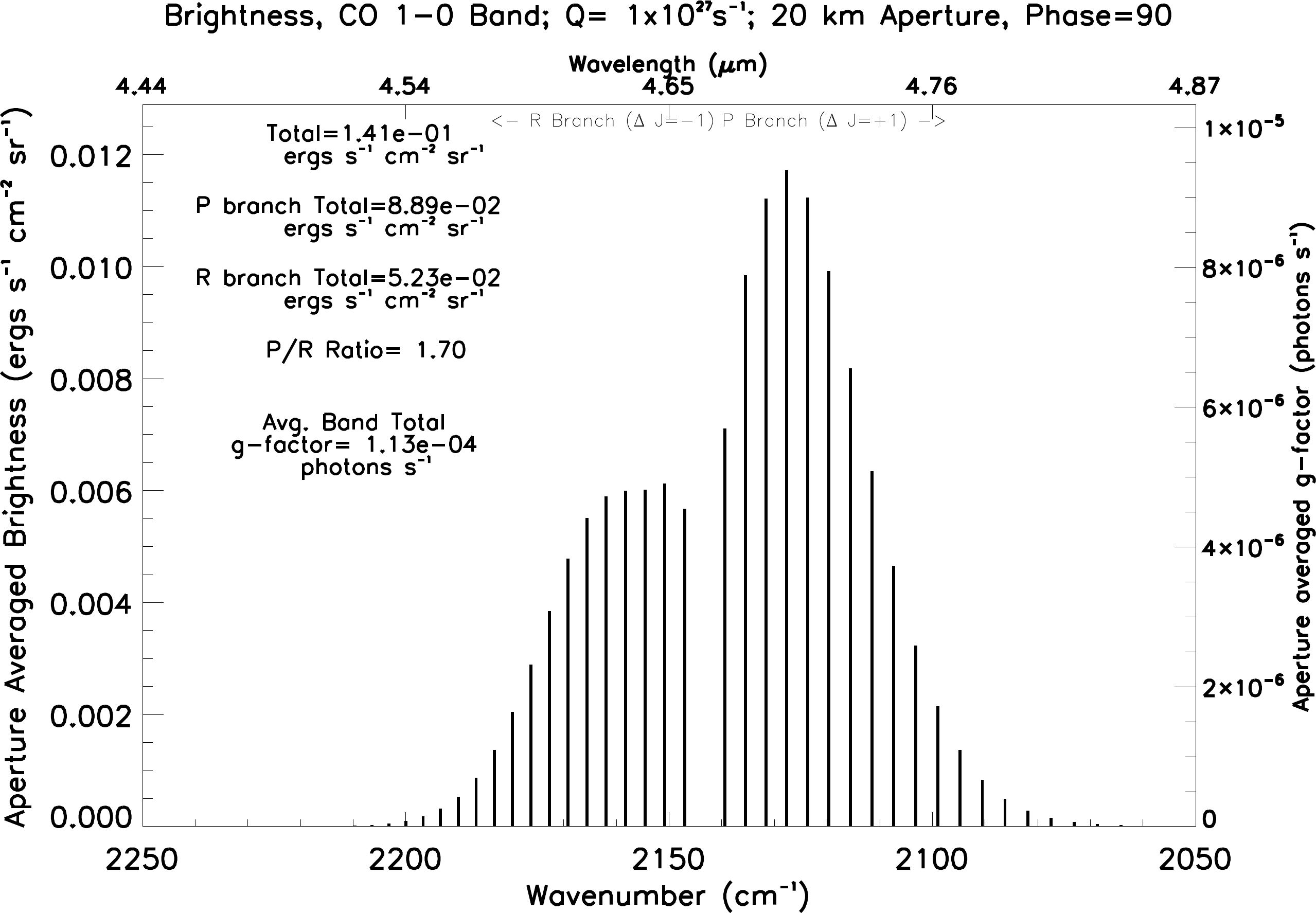}}
	\vspace{0.0in}
\subfigure[For $Q_{CO} = 10^{27}$ s$^{-1}$. Aperture = 100km.]{\label{aperture-R50-Q27}
          \includegraphics[width=0.35\textwidth,height=0.45\textwidth,angle = 90]{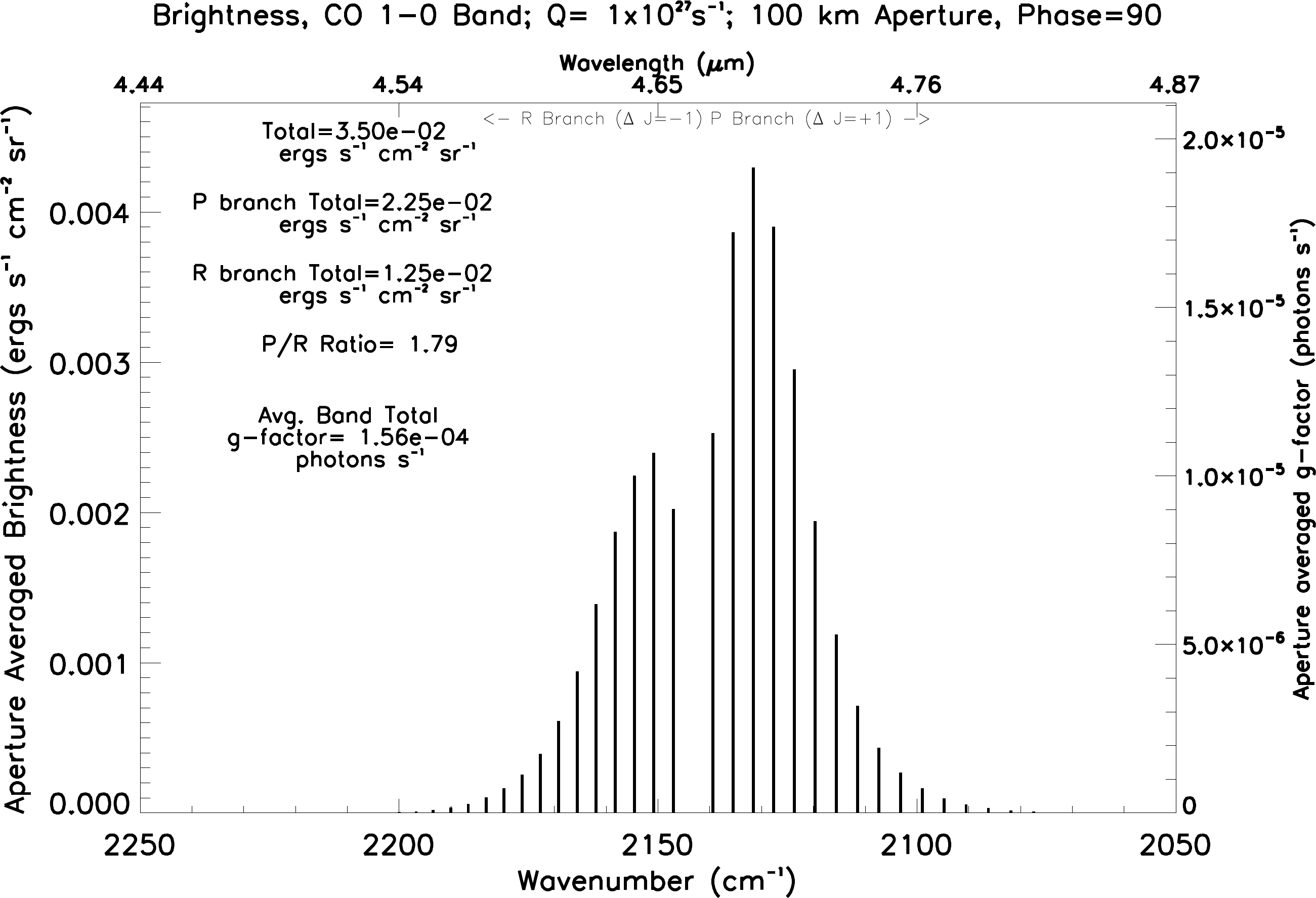}}
	\vspace{0.0in}
\subfigure[For $Q_{CO} = 10^{27}$ s$^{-1}$. Aperture = 200km.]{\label{aperture-R100-Q27}
          \includegraphics[width=0.35\textwidth,height=0.45\textwidth,angle = 90]{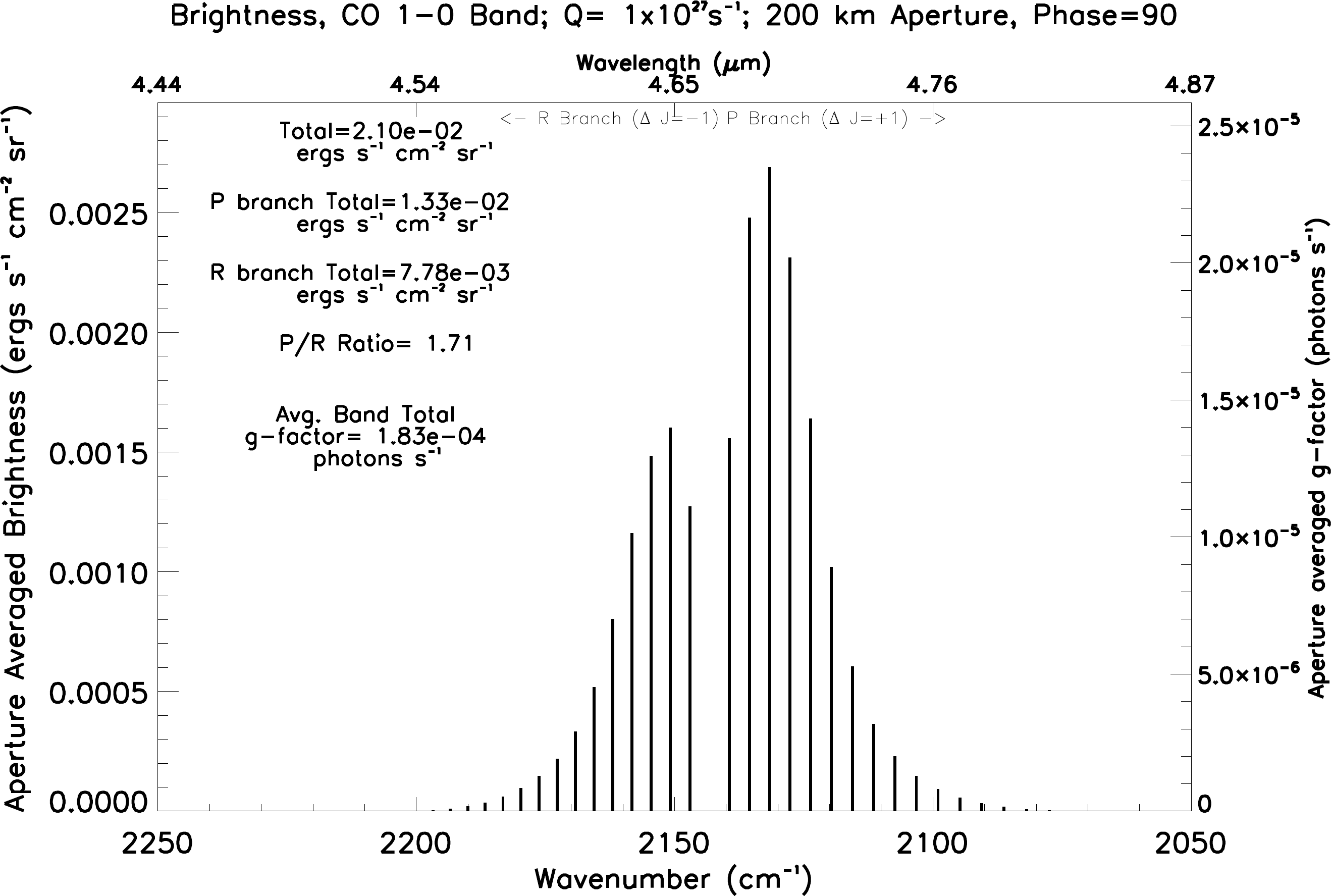}}
	\vspace{0.0in}
\subfigure[For $Q_{CO} = 10^{27}$ s$^{-1}$. Aperture = 2,000km.]{\label{aperture-R1000-Q27}
          \includegraphics[width=0.35\textwidth,height=0.45\textwidth,angle = 90]{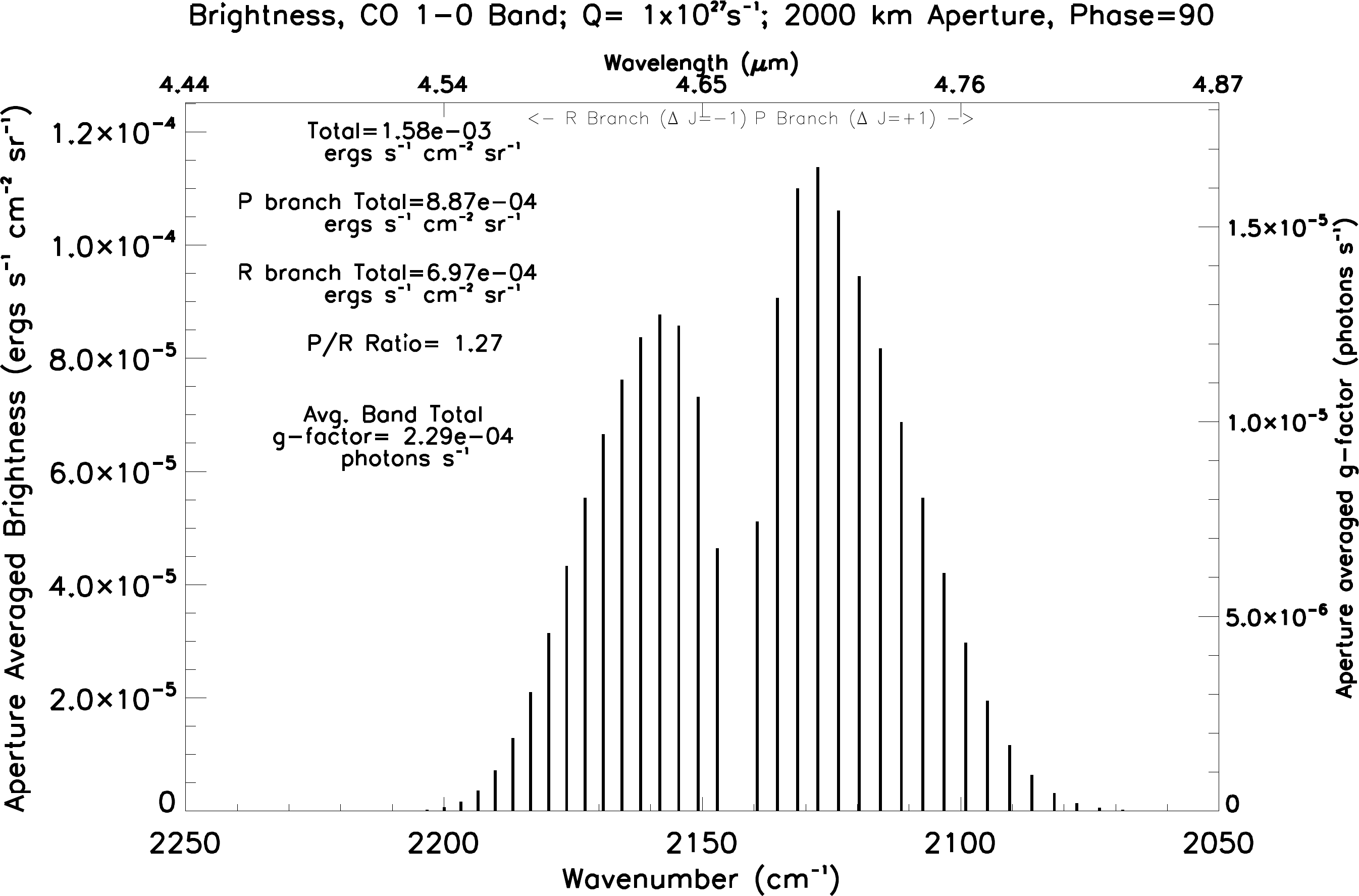}}
	\vspace{0.0in}
\subfigure[For $Q_{CO} = 10^{27}$ s$^{-1}$. Aperture = 20,000km.]{\label{aperture-R10000-Q27}
          \includegraphics[width=0.35\textwidth,height=0.45\textwidth,angle = 90]{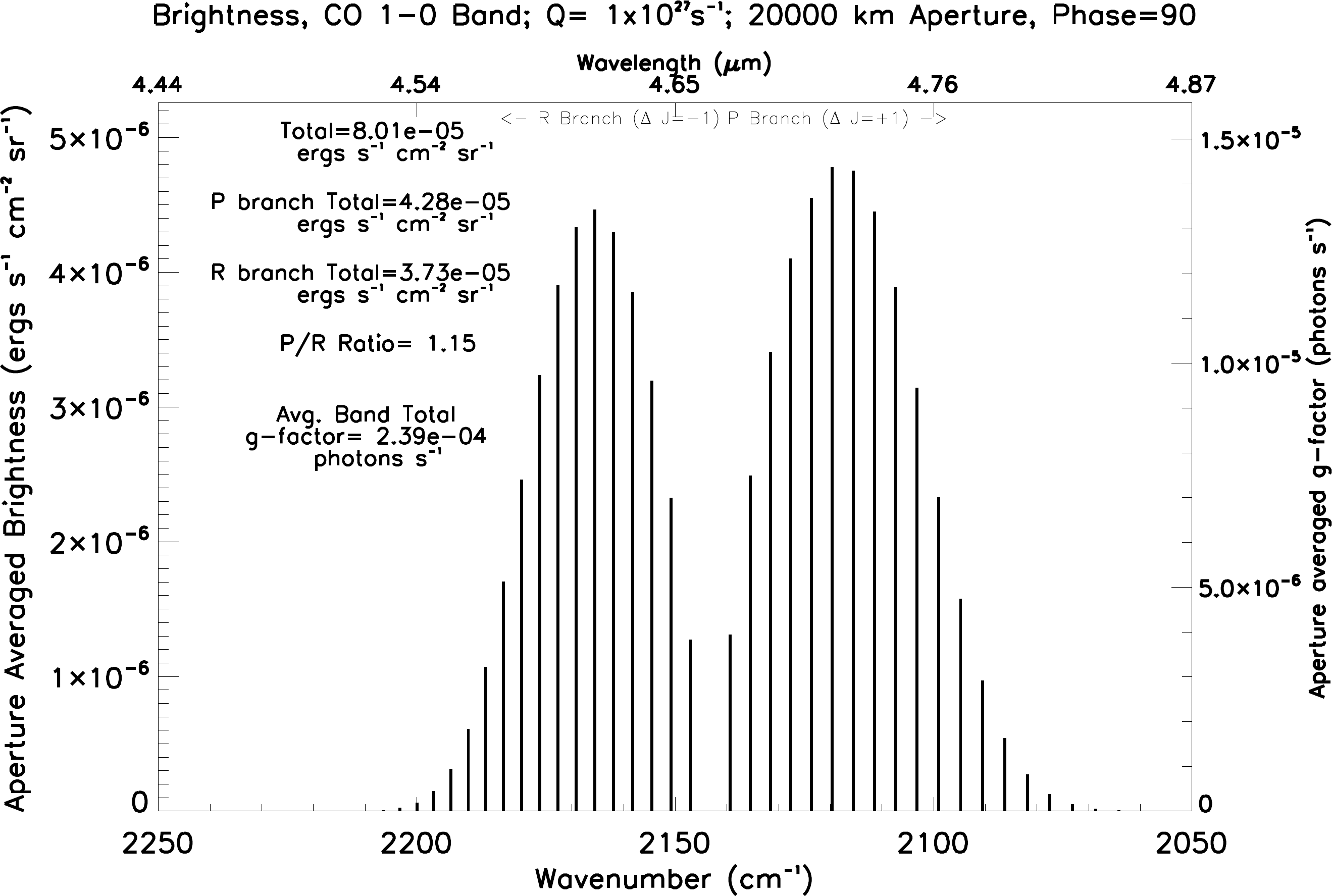}}
	\vspace{0.0in}
\subfigure[For $Q_{CO} = 10^{27}$ s$^{-1}$. Aperture = 200,000km.]{\label{aperture-R100000-Q27}
          \includegraphics[width=0.35\textwidth,height=0.45\textwidth,angle = 90]{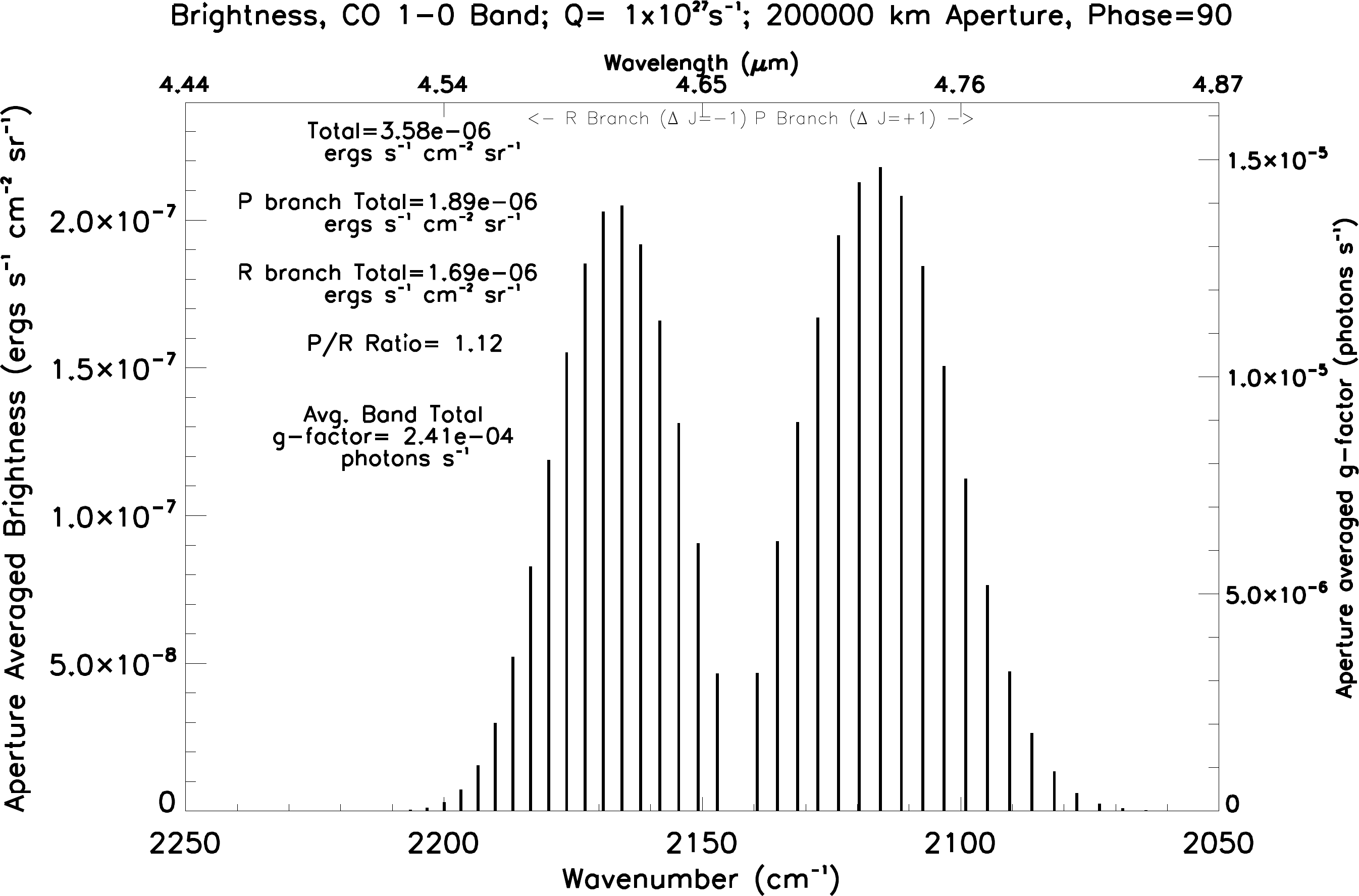}}
	\vspace{0.0in}
\caption{Aperture integrated spectra for $Q_{CO} = 10^{27}$ s$^{-1}$. Left side y-axis is aperture averaged brightness. Right y-axis is effective (line) g-factor (brightness/column density). Totals are indicated on each graph.\label{Q27spectra}}
\end{figure*}

\begin{figure*}[!ht]
\centering
\subfigure[For $Q_{CO} = 10^{28}$ s$^{-1}$. Aperture = 20km.]{\label{aperture-R10-Q28}
          \includegraphics[width=0.35\textwidth,height=0.45\textwidth,angle = 90]{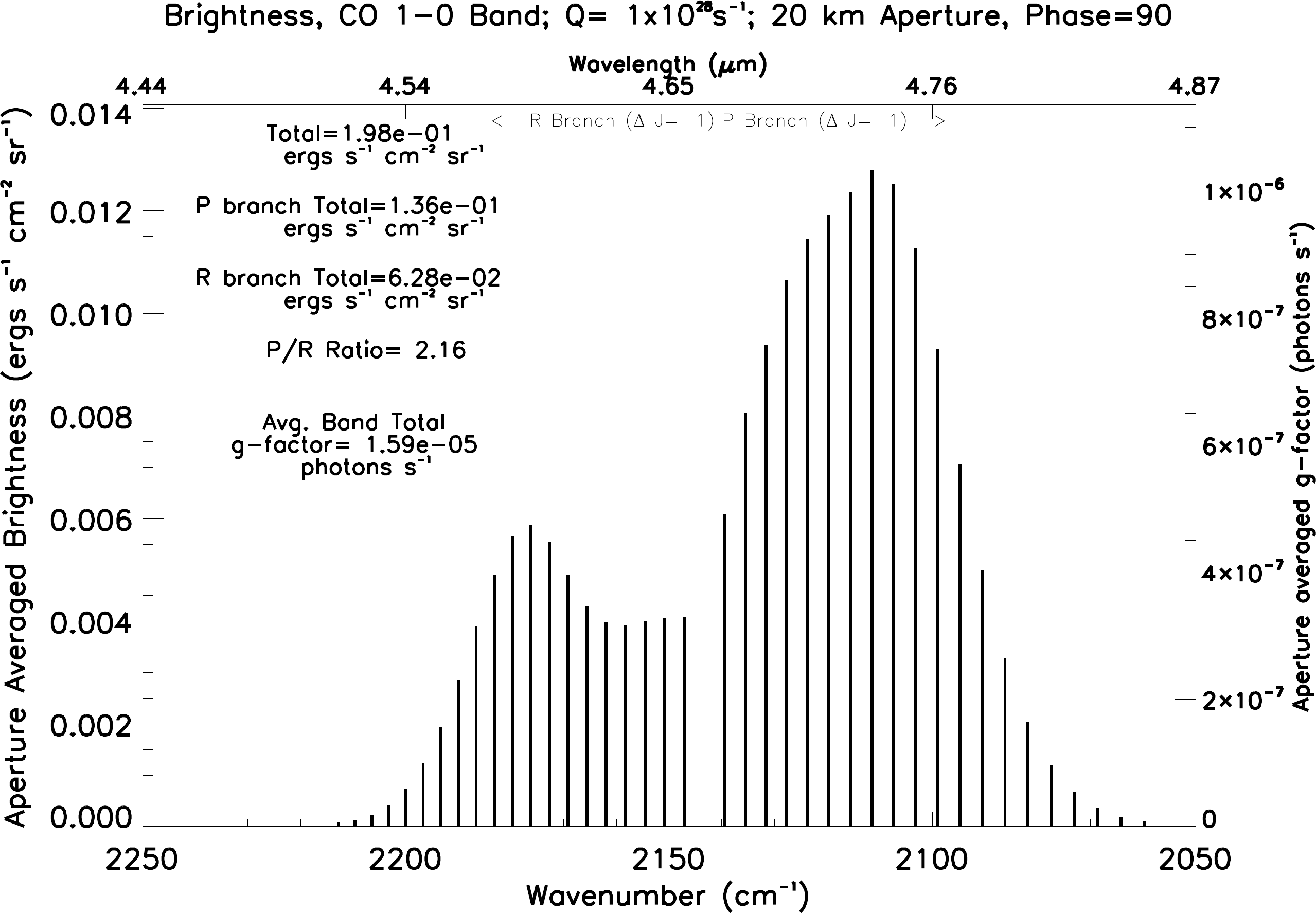}}
	\vspace{0.0in}
\subfigure[For $Q_{CO} = 10^{28}$ s$^{-1}$. Aperture = 100km.]{\label{aperture-R50-Q28}
          \includegraphics[width=0.35\textwidth,height=0.45\textwidth,angle = 90]{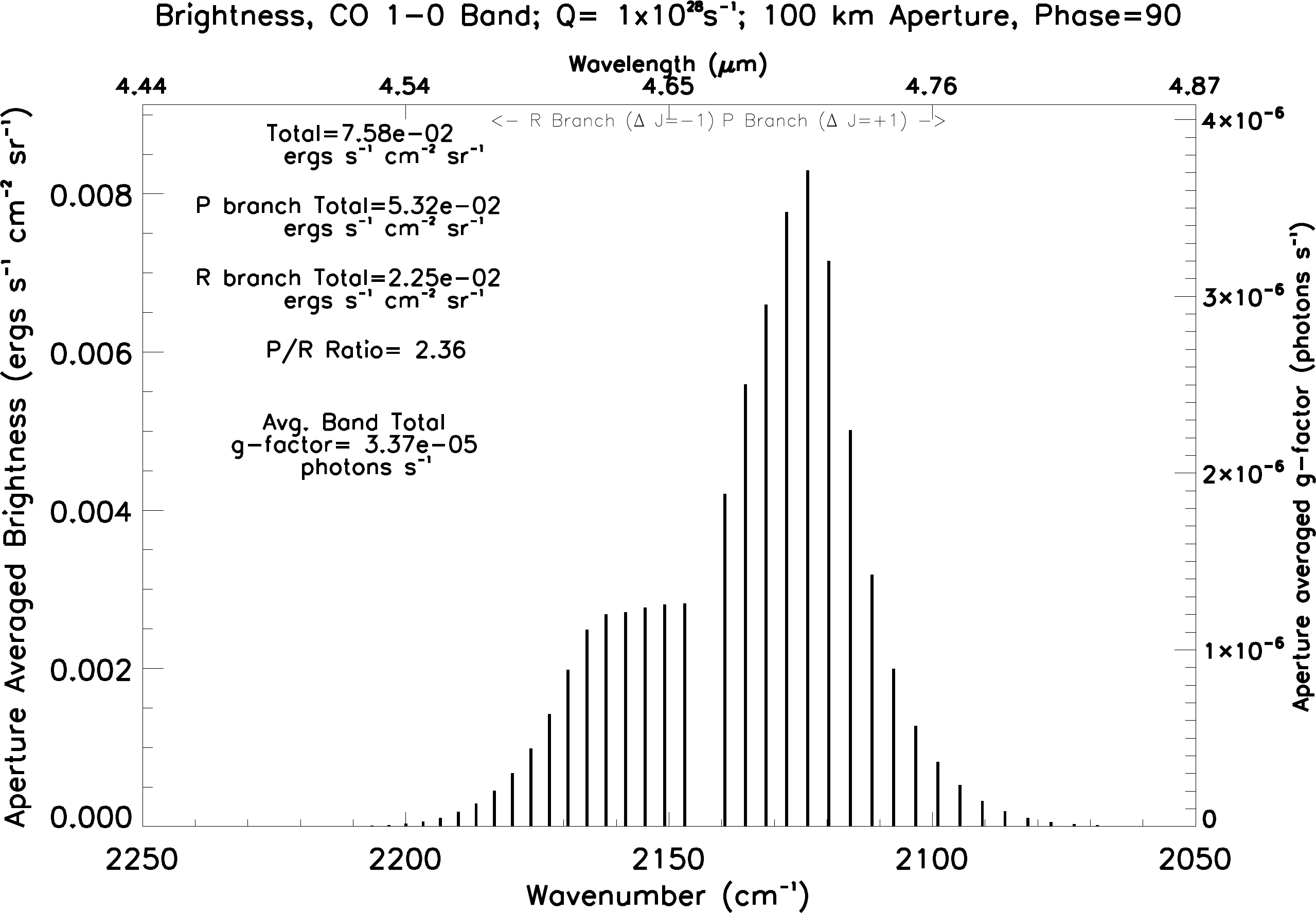}}
	\vspace{0.0in}
\subfigure[For $Q_{CO} = 10^{28}$ s$^{-1}$. Aperture = 200km.]{\label{aperture-R100-Q28}
          \includegraphics[width=0.35\textwidth,height=0.45\textwidth,angle = 90]{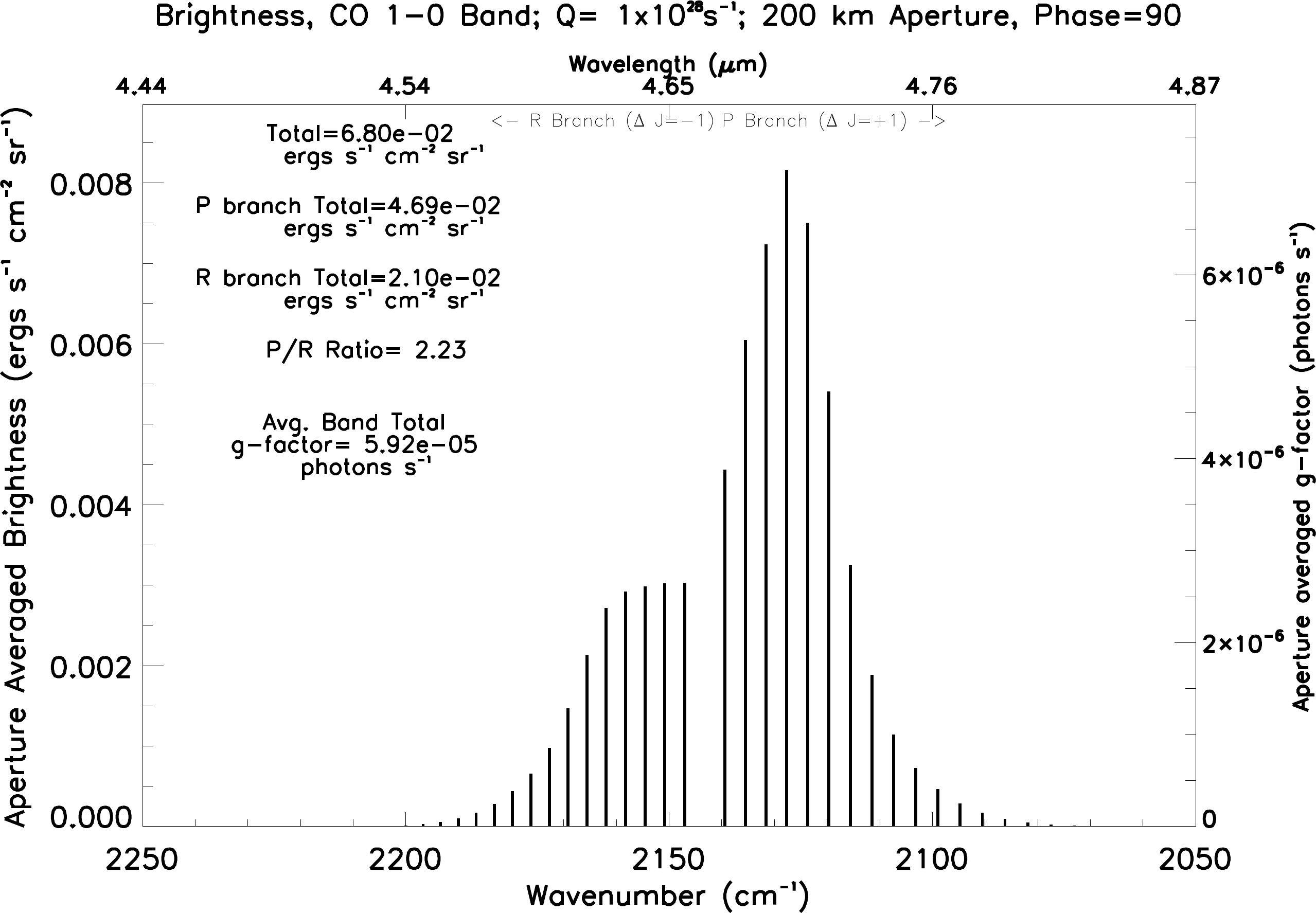}}
	\vspace{0.0in}
\subfigure[For $Q_{CO} = 10^{28}$ s$^{-1}$. Aperture = 2,000km.]{\label{aperture-R1000-Q28}
          \includegraphics[width=0.35\textwidth,height=0.45\textwidth,angle = 90]{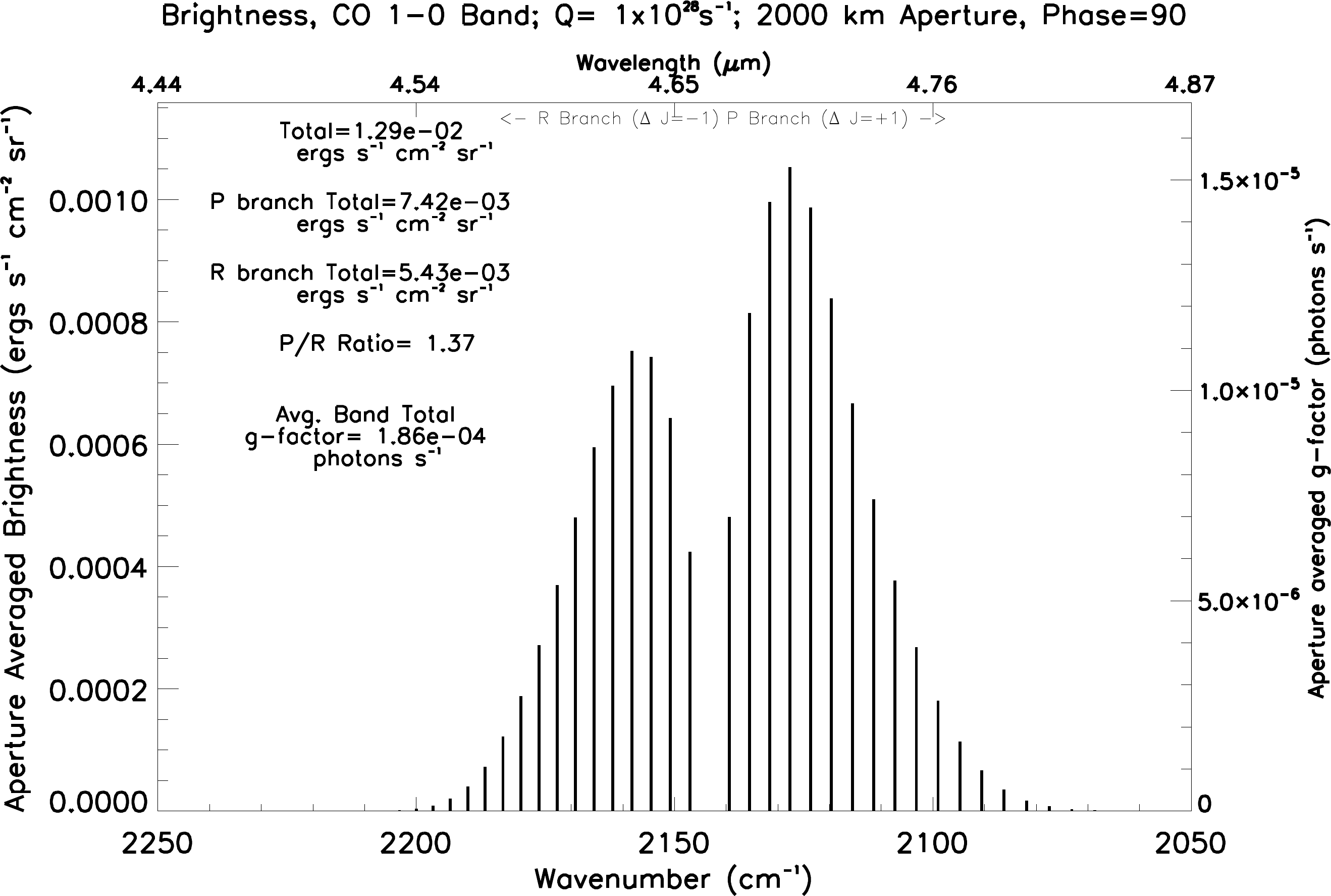}}
	\vspace{0.0in}
\subfigure[For $Q_{CO} = 10^{28}$ s$^{-1}$. Aperture = 20,000km.]{\label{aperture-R10000-Q28}
          \includegraphics[width=0.35\textwidth,height=0.45\textwidth,angle = 90]{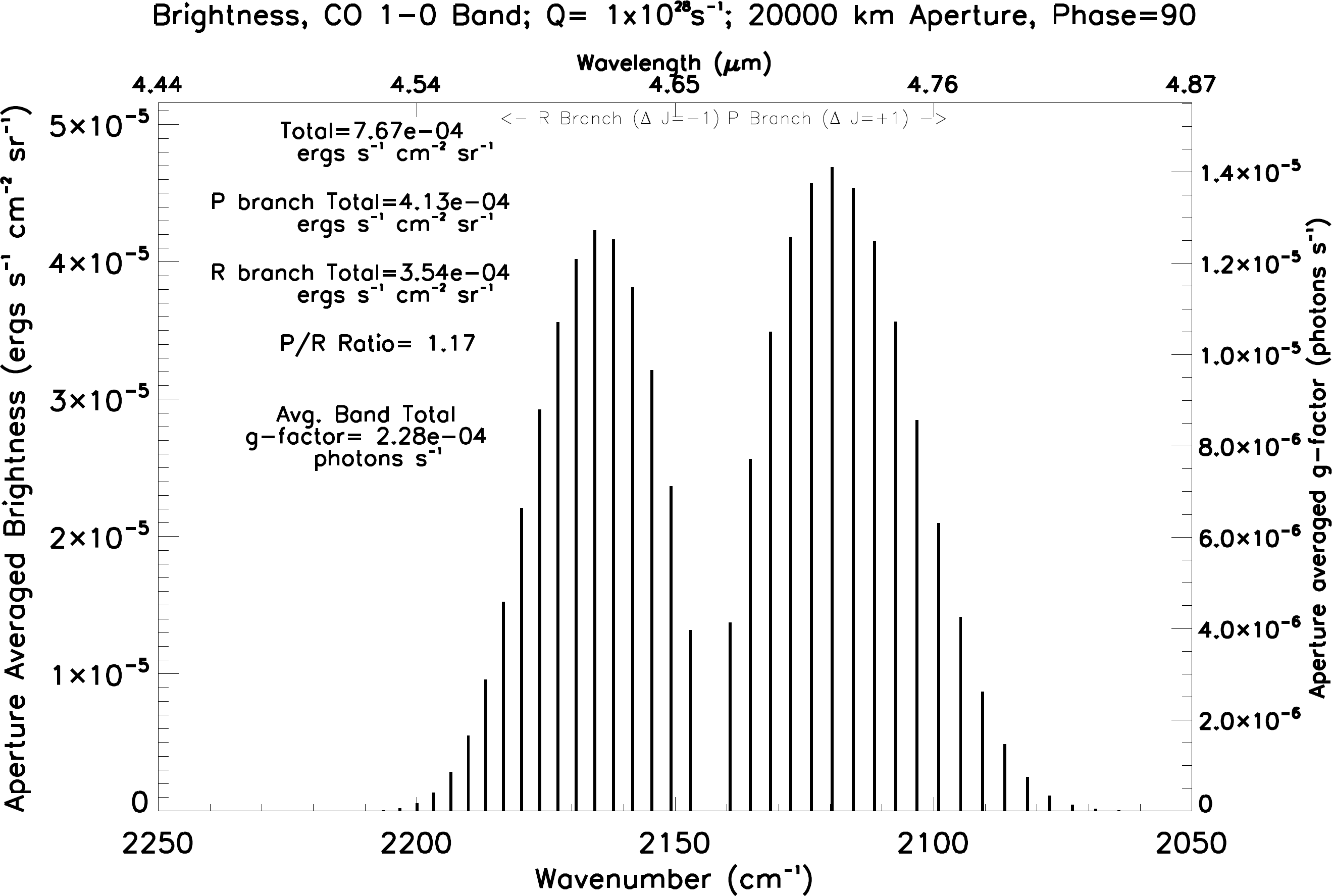}}
	\vspace{0.0in}
\subfigure[For $Q_{CO} = 10^{28}$ s$^{-1}$. Aperture = 200,000km.]{\label{aperture-R100000-Q28}
          \includegraphics[width=0.35\textwidth,height=0.45\textwidth,angle = 90]{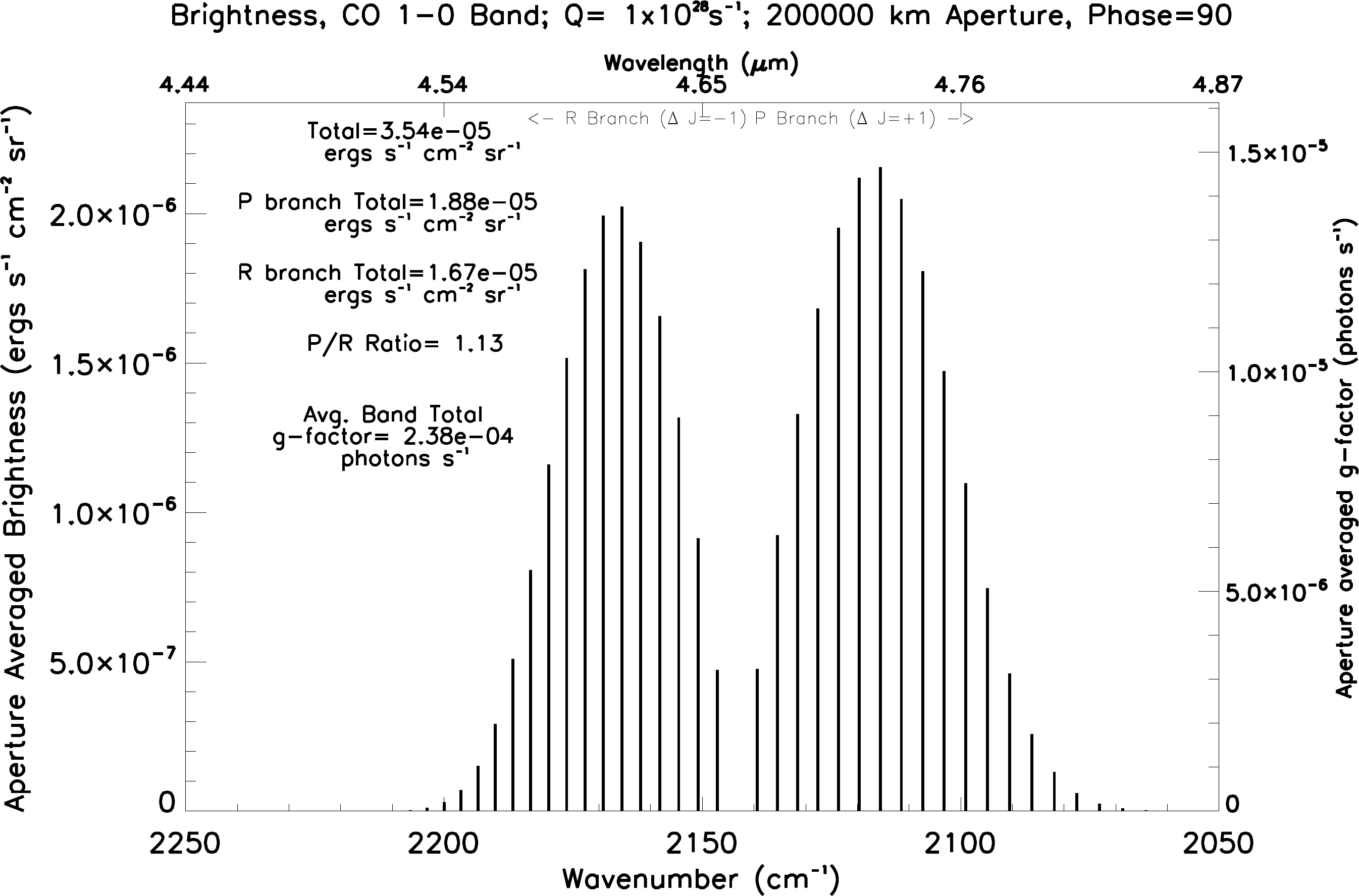}}
	\vspace{0.0in}
\caption{Aperture integrated spectra for $Q_{CO} = 10^{28}$ s$^{-1}$. Left side y-axis is aperture averaged brightness. Right y-axis is effective (line) g-factor (brightness/column density). Totals are indicated on each graph.\label{Q28spectra}}
\end{figure*}

Band shape for apertures including the outer coma (approximately $10^4 < R_{ap} < 10^5$ km) does not change significantly for different production rates. The total brightness for this regime increases approximately in linear proportion to production rate. This is due to spectra with such large aperture sizes being dominated by the fluorescence dominated optically thin outer coma with optical depth effects playing a minimal role. (But not entirely non-existent: note the small, $\lesssim 6\%$ reduction in g-factor with higher production rate for $R_{ap} = 2\times 10^4$ km.)

In the inner coma, however, optical depth effects can be very striking. 
The ``thickest'' spectra (for higher production rates and smaller $R_{ap}$) have remarkably altered band shapes from the optically thin spectra.

First, however, a word about changes that are {\it not} specifically caused by optical depth effects. It is clear that there is much variation, even within a given production rate, from large to small aperture sizes. Not all of this is due to optical depth effects. Even in our optically thin case, the band shape changes noticably in breadth. This occurs primarily due to the temperature profile.
We have used a fairly simplified profile, which can be scaled to a surface temperature parameter, but does not vary much otherwise between different model cases. This provides a straightforward ``control'' for this aspect of spectral change with aperture size.

In the innermost coma near the nucleus, the temperature is quite warm ($\sim$100 - 200K), which leads to a broader band in the 10 km spectrum. The coma gas cools to a minimum ($\sim$20K) around 100-200 km out from the nucleus, which produces a much narrowed band. Since the Einstein A coefficient for the lowest J lines is higher than for the lines in the ``wings'' of the band, the cold temperature also increases the g-factor, even in optically thick regions. At larger radii, the temperature rises again, but becomes less significant since the coma gets less dense and tends towards fluorescent equilibrium. Between these regimes, in a ``transition region,'' there are still optical depth effects, which can be more easily isolated as g-factors are less temperature controlled.

Temperature is also a factor in determining doppler broadening and line width, which is proprtional to $T^{1/2}$, so the ratio between line widths for the coldest and the warmest regions of the coma are about 2-3, for a given wavenumber.
This may lead to temperature playing a significant role in the optical thickness of the coma to incident solar radiation.

Temperature effects notwithstanding, the spectra from the denser near-nucleus regions of a coma show optical depth effects in several aspects. In addition to the total brightness no longer increasing linearly with production rate (and a corresponding reduction of g-factors), energy is also dramatically shifted between lines within the band. 

The notable shifting of flux from R branch to P branch (evident in many of the thicker spectra), and to lower wavenumbers in both branches (as is most evident in the $R_{ap} = $100 \& 200 km spectra for $Q=10^{28}$ s$^{-1}$), are very noticeable optical depth effects. (See Sahai \& Wannier 1985, for an analytical discussion of similar effects.) 

This effect appears due to the branching ratio of a given pair of P and R branch lines originating in the same upper level, which generally (slightly) favors emission in the P branch line. In optically thick cases, repeated absorption and emission of photons leads to a cumulative effect which favors the P branch over the R branch much more than in optically thin conditions (where it is probable that any emitted photon will not be re-absorbed before escaping the coma).

Similarly, flux is ``pushed'' outwards in the branches, and more so in the P branch due to combination with the above effect. This is due to the lines closer to the center of the band becoming optically thick before those in the wings (both due to their higher Einstein coefficients and generally being more populated.) Flux initially emitted in lines that are optically thick will through repeated absorption and emission be forced out into lines that are less thick.

\subsection{The P/R Ratio: A Useful Heuristic of Optical Depth}

As seen above, the P branch total brightness and the R branch total brightness vary with respect to each other over different optical depths (as well as other factors, such as temperature distribution.) The ratio of the sums of P/R branches' brightnesses can be useful to alert an observer (or anyone analyzing observations) that they must in a given case beware of, and if possible account for, optical depth effects. 

This alone, would not be sufficient, as temperatures along a given line of sight are also a significant factor in controlling the P/R ratio, in the collisionally dominated inner coma.
Colder population distributions will emit more in the lower lines (in both branches) for which the ratio of P/R for each pair of lines originating in the same upper state is greater. 
Also, $d\tau$ will effectively vary inversely with linewidth, other factors being equal.

Use of a model like ours can show where the P/R ratio is large due to temperature and where (its excess beyond that value is) due to optical depth. In our optically thin, $Q_{CO} = 10^{26}$ s$^{-1}$, model the P/R ratio does not exceed $\sim$1.44, even for aperture sizes dominated by the coldest portion of the coma. (Note, however, that this is an aperture {\it averaged} value. In Fig. ~\ref{PR_Q26} below, the peak value is slightly higher, $\sim$1.5.)  However, the ratio for corresponding aperture sizes in the $Q_{CO} = 10^{27}$ s$^{-1}$ and $Q_{CO} = 10^{28}$ s$^{-1}$ cases is $\sim$1.8 and $\sim$2.4, respectively. Furthermore, in the thickest case modeled, even the spectrum with aperture size of 2000 km has a ratio of  $\sim$1.4. Note that in all cases the $2\times10^{5}$ km aperture, which is dominated by the outer coma in fluorescent equilibrium, has a ratio of $\sim$1.12. All of this indicates that a P/R ratio in excess of  $\sim$1.4$\sim$1.5 is a warning sign of optical depth effects involved.

\subsection{Further Discussion}

While it would be ideal to be able to derive a simple correction factor from the P/R ratio in such cases, alas, it is not exactly possible. However, a rough estimate of the degree of optical depth effects can be derived.

\begin{figure}[h]
\centerline{\includegraphics[width=0.4\textwidth, angle=90]{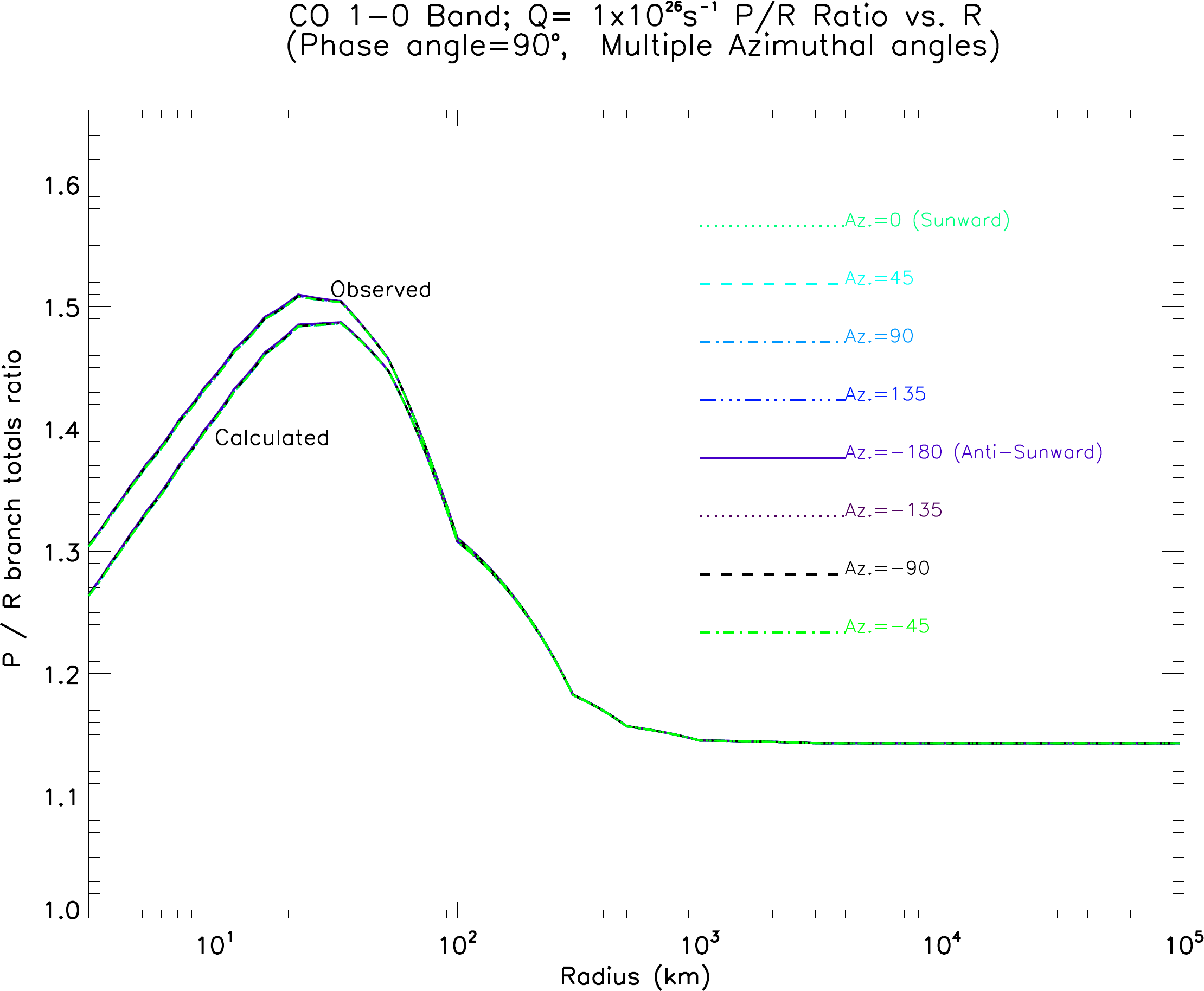}}
\caption{For $Q_{CO} = 10^{26} s^{-1}$. Ratio of P branch vs. R branch total Brightness vs. R (impact paramater) for Phase angle = 90\degree and for multiple Azimuthal angles. Profiles of azimuthal angles show negligible variation for this case and overlap, appearing indistinguishable. (Profiles are indicated by color coding in the online version). \label{PR_Q26}}
\end{figure}

\begin{figure}[h]
\centerline{\includegraphics[width=0.4\textwidth, angle=90]{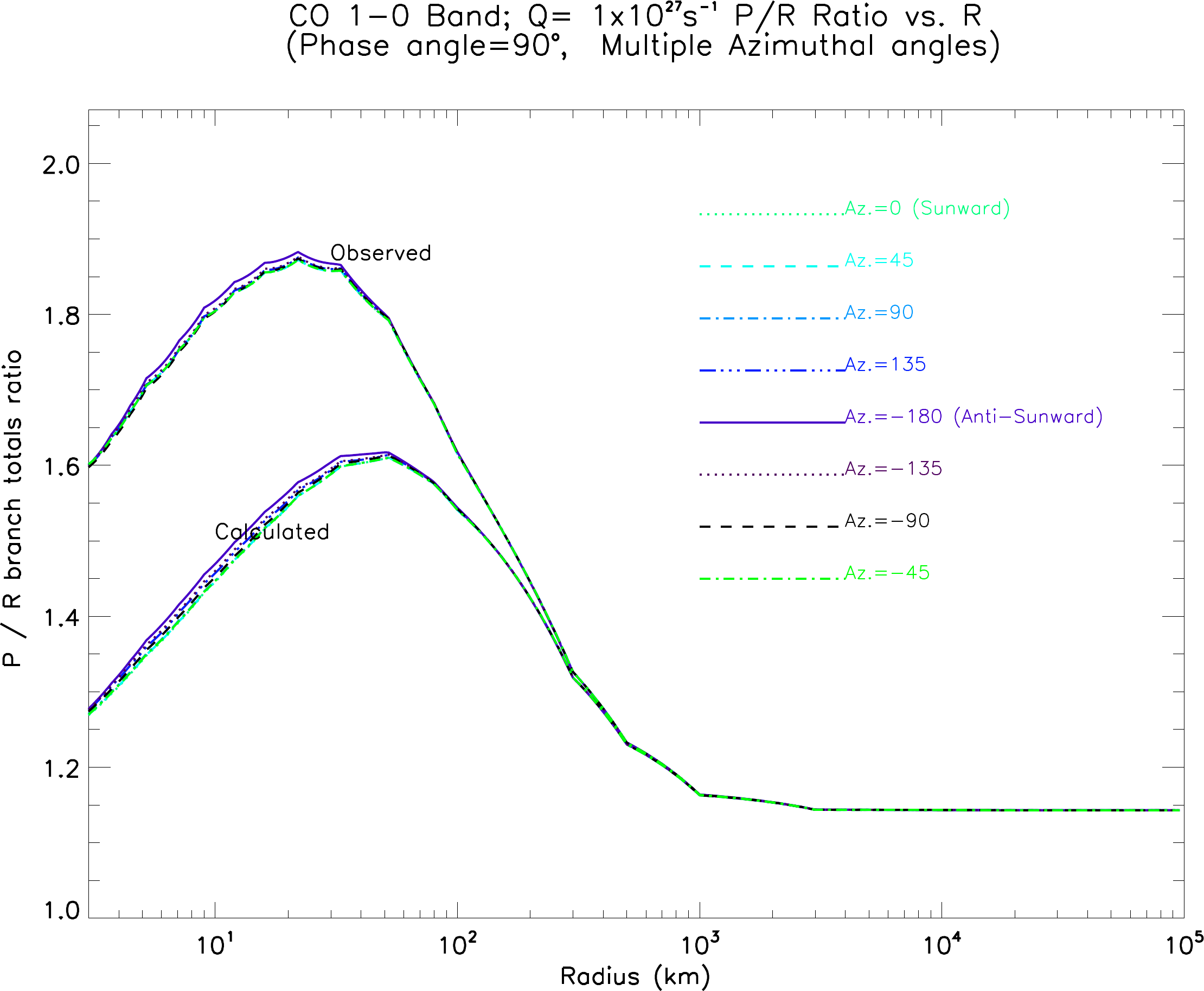}}
\caption{For  $Q_{CO} = 10^{27} s^{-1}$. Ratio of P branch vs. R branch total Brightness vs. R (impact paramater) for Phase angle = 90\degree and for multiple Azimuthal angles.  Profiles of azimuthal angles show minimal variation for this case and overlap, appearing {\it nearly} indistinguishable, except inwards of $\sim$40-50 km. (Profiles are indicated by color coding in the online version). \label{PR_Q27}}
\end{figure}

\begin{figure}[h]
\centerline{\includegraphics[width=0.4\textwidth, angle=90]{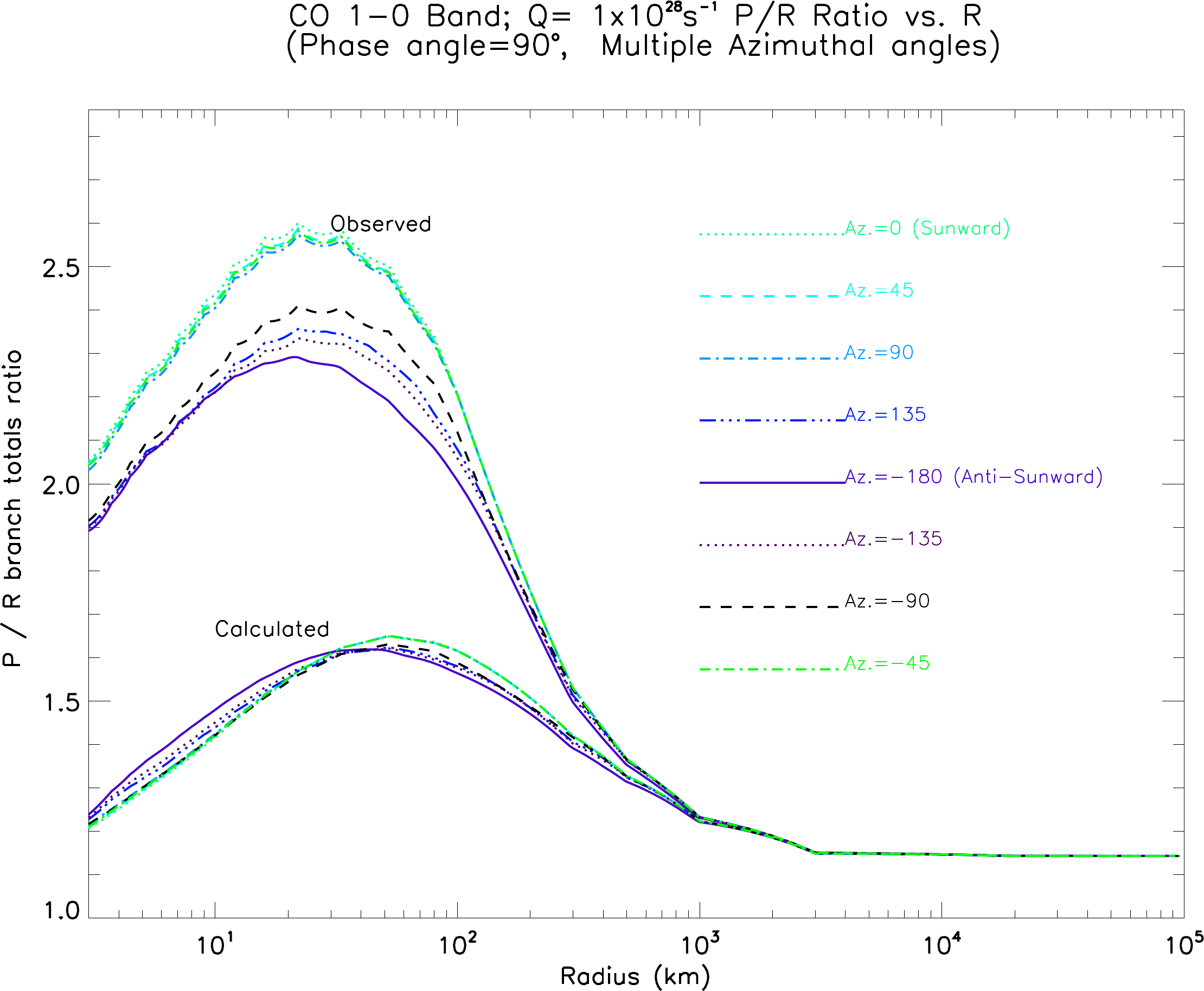}}
\caption{For  $Q_{CO} = 10^{28} s^{-1}$. Ratio of P branch vs. R branch total Brightness vs. R (impact paramater) for Phase angle = 90\degree and for multiple Azimuthal angles. Profiles of azimuthal angles show some variation in this case, but those for 0\degree, $\pm$45\degree, and 90\degree overlap each other entirely in the calculated profiles and almost entirely in the observed. (Profiles are indicated by color coding in the online version). \label{PR_Q28}}
\end{figure}

To do so, we create radial profiles of the P/R ratio, for both the observed emergent flux/brightness and the calculated value based on underlying populations without attenuation of emergent light, as shown in Figs. ~\ref{PR_Q26}, ~\ref{PR_Q27}, and ~\ref{PR_Q28}. By cross-referencing the observed P/R ratio for a given radial distance with the corresponding g-factor in Figs. ~\ref{fluxVsRphase90multiAzQ26},  ~\ref{fluxVsRphase90multiAzQ27}, and  ~\ref{fluxVsRphase90multiAzQ28} one can ascertain the ``real'' g-factor to use to calculate a correct column density from the observed flux.

This heuristic is, of course, limited in use to to the carbon monoxide $X^{1} \Sigma^{+}$ band. Other spectra with P and R branches will have their own ratios, which can be derived by similar modeling. More complicated spectra may also, but such ratios would be more complicated to find than for cases with a simple two-branch structure.

\section{Next Steps}

We plan to provide further useful results in forthcoming papers (currently in preparation) dealing with similar modeling to that presented here for H$_2$O and CO$_2$, as well as a more in depth treatment of CO alongside those species. We will present model results including morphology and comparison with the in-situ spectral observations of the Deep Impact and EPOXI missions.
The model and code have already been implemented to produce those results.

Further work, not yet implemented, will include several planned improvements to our model. 
A more accurate treatment of radial velocites and doppler shifts in lines due to them is a highly desirable improvement.
(At present, only a thermal doppler profile is used for calculating absorption of solar radiation.) We neglected a precise treatment until now primarily due to the nature of our originally intended problem. The near nucleus morphology which we intended to model is represented by cones expanding radially in one direction, a geometry that is likely to reduce doppler effects. However, when modeling the whole coma as a sphere, as in the results presented here, this assumption is no longer valid. 
Furthermore, in as much as we are primarily looking at the overall {\it band} shape as opposed to individual line shapes, the effect of doppler shifts on the spectra is expected to be less than if one were modeling specific lines (as is often the case for other spectral regimes). However, the effect of this neglect would be to increase the optical thickness, and therefore our results may be over-estimating optical depth effects. If so, the effects we describe would still be observable, but for higher production rates than those modeled.

In addition to more accurate radial velocities, more flexible radial temperature and density profiles are planned, so as to be able to model deviations from a very simple Haser model. (e.g. Volatiles produced from icy grains or large chunks and not solely from the nucleus' surface, as per A'Hearn, et al. 2011, and/or photodissociation of molecules, and corresponding creation of daughter species.)

Computational limits are currently a limiting factor in how optically thick and how refined (in terms of granularity of conditions, in that more variation requires more regions) the modeled cases can be. As of now, the maximum production rates we can deal with are on the order of $Q=10^{28}$ s$^{-1}$, and somewhat less for CO$_2$ than the other molecules. We are planning to address these concerns with algorithmic improvements. Running the code on faster and more powerful computers is also a possibility.

\section{Conclusions}

We have demonstrated our model's usefulness in understanding emission spectra of cometary comae. There are several possible effects of optical depth that could lead observers to mistaken conclusions regarding the calculated abundances, or other characteristics, of species of interest.
The moral of the story: Ignore radiative transfer and optical depth effects at your own peril!

Although designed specifically with comets in mind, our model and code are versatile enough to be used in other radiative transfer problems as well.  Parameters that define a specific comet model or other problem, including molecule of interest, size of nucleus and radial shells, production rate, morphology (if any), incident radiation, etc. are all fairly flexible.
Thus our adaptation of Coupled Escape Probability to an asymmetric spherical situation has created a very useful tool for modeling cometary spectra, as well as other spherical astrophysical phenomena.

\begin{appendix}

\section{Algorithm Implementation \& Technical Details}

We have coded the aglorithm described above in the C++ language, using numerous functions from Press, et al. 1992, primarily to implement numerical integration of functions (with {\bf odeint} and {\bf stifbs} and associated functions) and solution of N-dimensional non-linear matrices with Newton's Method (using {\bf newt}, and associated functions). The bulk of the coding, which implements the radiative transfer algorithm in spherical geometry is our own.

A major practical limitation of our algorithm is the matrix size; since Newton's method requires (repeated) $O(N^3)$ matrix solving operations (for which {\bf newt} uses the brute force approach of {\bf ludcmp} and {\bf lubksb}), the algorithm can get prohibitively slow for large matrices. 
For example, on an Intel Core computer (with CPU speed of 2.9 GHz and 7.6 GB of memory) running Scientific LINUX 6.3 (Carbon), when N $\gtrsim$ 15,000, a solution may take one or more days. For greater sizes, even a week. The matrix size equals the number of molecular levels used times the number of regions. The molecule (and band/s) being modeled determines the first value, with the latter value demanding to be increased with greater optical depths and production rates. Depending on the species of interest, the maximum practical production rates we can currently manage on such a system are of $O(10^{28}) s^{-1}$.
The most simple workaround is to use a more powerful computer. Algorithms for solving sparse matrices faster than the above functions can also be used, and we have begun to explore this option.

In implementing the CEP method, and our adaptation to spherical geometry, we have created C++ classes for diatomic and triatomic molecules. The object oriented programming style of C++ lends itself ideally to being able to switch molecules easily. The {\bf Diatomic} molecule class and its subclasses (currently implemented for {\bf CO} and {\bf SiO}) calculate energies and Einstein coefficients based on constants (taken primarily from Krupenie 1966, for CO) that are included in the code. The {\bf Triatomic} class, which can actually be used for other polyatomic molecules as well (or the aforementioned diatomic molecules themselves, for that matter), must be provided with energies and coefficients from some outside source in formatted input files. We used the HITRAN database (see Rothman, et al. 1998) to supply these values for CO$_2$ and H$_2$O. This approach has the versatility to handle many other molecules with a minimal effort of ``data massaging'' to get the data into the proper format.
We have also created a large number of classes which encapsulate the spherical geometry and the attendant calculations. These are all ``controlled'' by the {\bf Comet} class that reads parameters for a given case from a ``comet definition file'' (a text file of key/value pairs, essentially) and, using the above classes, sets up and runs the model for a given case.

Although designed specifically with comets in mind, our code is versatile enough to be used in other spherical radiative transfer problems as well. Straightforward input files describe all the required and optional parameters for a specific comet model or other problem, including molecule of interest, size of nucleus and radial shells, production rate, morphology (if any), incident radiation, etc. 

The C++ code outputs a file containing a point-by-point line-by-line spectral mapping, as described above, for one or more specified viewing orientations. This data can then be presented in multiple formats as described above in Section 3. We have implemented this data presentation portion of the model using various short IDL programs which we have developed, each specifically for the purpose of producing a given type of graphical output.

\end{appendix}

\acknowledgements

{\bf Acknowledgements} We gratefully thank Profs. J. P. Harrington and D. C. Richardson and Dr. L. Feaga for their invaluable advice in the course of this work. 
We would also like to thank Prof. M. Elitzur for his encouragement.
This work was supported by NASA's Discovery Program contract NNM07AA99C to the University of Maryland.

\renewcommand{\bibnumfmt}[1]{{}}

\citeindexfalse

 \end{document}